\documentclass[12pt,a4paper]{article}
\usepackage{amsmath,caption,bbm,longtable,paralist}
\usepackage[top=1.3in, bottom=1.3in, left=1.1in, right=1.1in]{geometry}
\usepackage[skip=1pt,font=small]{caption}

\DeclareCaptionStyle{italic}[justification=centering]{labelfont={bf},textfont={it},labelsep=colon}
\usepackage{graphicx,psfrag,epsf}
\usepackage{enumerate}
\usepackage{natbib}
\usepackage{url} 
\usepackage{booktabs, subfig, bm, paralist,mathpazo,tikz,todonotes,longtable,microtype,dsfont,rotating} 
\usepackage[pdftex,colorlinks=true]{hyperref}
\definecolor{darkblue}{rgb}{0,0,.6}
\hypersetup{citecolor=darkblue,linkcolor=darkblue,urlcolor=darkblue}

\newcommand{\blind}{0}

\graphicspath{{plots/}}

\newsavebox\CBox
\def\textBF#1{\sbox\CBox{#1}\resizebox{\wd\CBox}{\ht\CBox}{\textbf{#1}}}

\definecolor{a0}{rgb}{0.0, 0.5, 0.0}
\definecolor{bistre}{rgb}{0.24, 0.17, 0.12}
\definecolor{amethyst}{rgb}{0.6, 0.4, 0.8}
\definecolor{blue-violet}{rgb}{0.54, 0.17, 0.89}
\definecolor{Rcolor}{RGB}{150,160,190}
\definecolor{blush}{rgb}{0.87, 0.36, 0.51}
\definecolor{brightturquoise}{rgb}{0.03, 0.91, 0.87}
\definecolor{burntorange}{rgb}{0.8, 0.33, 0.0}

\date{\today}
\AtBeginDocument{}

\begin{document}

\def\spacingset#1{\renewcommand{\baselinestretch}
{#1}\small\normalsize} \spacingset{1}

\if0\blind
{
  \title{\bf Estimation of a functional single index model with dependent errors and unknown error density}
  \author{Han Lin Shang\thanks{Postal address: Research School of Finance, Actuarial Studies and Statistics, Level 4, Building 26C, Australian National University, Kingsley Street, Acton Canberra, ACT 2601, Australia; Telephone: +61(2) 612 50535; Fax: +61(2) 612 50087; Email: hanlin.shang@anu.edu.au.}  \hspace{.2cm}\\
    Research School of Finance, Actuarial Studies and Statistics \\
    Australian National University \\
}
  \maketitle
} \fi

\if1\blind
{
  \bigskip
  \bigskip
  \bigskip
  \begin{center}
    {\LARGE\bf Title}
\end{center}
  \medskip
} \fi

\bigskip

\begin{abstract}
The problem of error density estimation for a functional single index model with dependent errors is studied. A Bayesian method is utilized to simultaneously estimate the bandwidths in the kernel-form error density and regression function, under an autoregressive error structure. For estimating both the regression function and error density, empirical studies show that the functional single index model gives improved estimation and prediction accuracies than any nonparametric functional regression considered. Furthermore, estimation of error density facilitates the construction of prediction interval for the response variable.
\end{abstract}

\\


\newpage
\spacingset{1.4}

\section{Introduction}\label{sec:intro}

Functional data analysis concerns the statistical analysis of data where at least one of the variables of interest is a function. Regression with functional data is arguably the most thoroughly researched topic within the broader literature on functional data analysis. It is common to classify functional regression models into three categories according to the role played by the functional data in each model: scalar-on-function regression, function-on-scalar regression, function-on-function regression. \cite{Morris15}, \cite{RGS+17} and \cite{FGG17} provide a detailed overview of linear and non-linear methods for functional regression. This article focuses on a non-linear semi-parametric model; namely, the functional single index model in~\eqref{eq:1}, applied to scalar-on-function regression.

The functional single index model is a semi-parametric regression model for estimating the relationship between the scalar response and the functional predictor. It assumes the existence of a latent univariate explanatory variable that explains the association with the response through a nonparametric regression model. In addition, the latent explanatory variable can be estimated by a parametric functional regression model, such as functional linear regression. The estimated regression coefficient is useful for interpretation, while the nonparametric regression of the functional single index allows us to capture the possible non-linear relationship between the function-valued predictor and the scalar-valued response, and in turn, improves the estimation and prediction accuracies of the regression function. Thus, it couples the advantages of both parametric and nonparametric regression models. Because of these advantages, it has received an increasing amount of attention in the functional regression literature \citep[e.g.,][]{FPV03, JS05, AFK+08, FPV11, CHM11, JW11, GV15, FJR15}.

Despite its rapid development in the estimation of functional single index models, error density estimation remains largely unexplored. However, the estimation of error density is important to understand residual behavior and to assess the adequacy of the error distribution assumption \citep[e.g.,][]{AV01}; error density estimation is vital for testing the symmetry of the residual distribution \citep[e.g.,][]{ND07}; and is useful for carrying out statistical inference, model validation and prediction \citep[e.g.,][]{Efromovich05}. Moreover, the estimation of error density is critical to the estimation of the density of the response variable \citep[e.g.,][]{EJ12}. In the area of financial risk management, a pivotal use of the estimated error density is to estimate value-at-risk or the expected shortfall for holding an asset. In such a model, any misspecification of the error density may produce an inaccurate estimate of potential risks. Thus, being able to estimate the error density is as important as being able to estimate the regression function.

Building upon the early work by \cite{Shang13, Shang14, Shang14b, Shang16} and \cite{ZKS14}, the unknown error density is approximated by a mixture of Gaussian densities with means being the individual residuals and variance as a constant parameter. The unknown error density has the form of a kernel density estimator of residuals, where the regression function, consisting of parametric and nonparametric components, can be estimated by functional principal component regression and univariate Nadaraya-Watson estimators. The advantage of the functional single index model is that it provides a data-driven way to determine the choice of semi-metric for measuring distances among functions, but its estimation and prediction accuracies depend crucially on the optimal selection of bandwidth parameter. We implement a Bayesian method for estimating bandwidths in the regression function and error density simultaneously. Differing from the existing literature where the errors are treated as independent and identically distributed (iid), we capture temporal dependency between errors through a stationary autoregressive model of order $p$ (AR($p$)) \citep[see, e.g.,][]{DZZ16}. Hence, we also model the autoregression parameter in our extended Bayesian bandwidth estimation method.

The rest of the paper is organized as follows. In Section~\ref{sec:2}, we introduce the functional single index model. The Bayesian bandwidth estimation method is described in Section~\ref{sec:2.3}. Because the functional single index model provides a data-driven way to estimate the choice of semi-metric among functions, its estimation and prediction accuracies of the regression function are likely to outperform nonparametric functional regression which also requires the additional optimal selection of semi-metric. Using a series of simulation studies in Section~\ref{sec:4}, we evaluate and compare the estimation accuracy of the regression function and error density, as well as the point forecast accuracy between the functional single index model and the nonparametric functional regression. With two spectroscopy data sets, the estimation and forecast accuracies of the regression function between the functional single index and the nonparametric functional regression models are evaluated and compared in Section~\ref{sec:5}. Section~\ref{sec:6} concludes the paper, along with some ideas on how the methodology can be further extended.

\section{Model and estimator}\label{sec:2}

\subsection{Functional nonparametric regression model}\label{sec:2.1}

We consider a random pair $(\mathcal{X}, y)$, where $y$ is real-valued and the functional random variable $\mathcal{X}$ is valued in some infinite dimensional semi-metric vector space $\left(\mathcal{F}, d(\cdot,\cdot)\right)$. Let $(\mathcal{X}_i, y_i)_{i=1,\dots,n}$ be a sample of pairs that are identically distributed as $(\mathcal{X}, y)$. We consider a simple nonparametric functional regression model with homoskedastic and correlated errors. Given a set of observations $(\mathcal{X}_i, y_i)$, the model can be expressed as
\begin{equation}
  y_i = m(X_i) + \varepsilon_i = m\left[\int_{\mathcal{I}} \mathcal{X}_i(t)\beta(t)dt\right] + \varepsilon_i, \qquad i=1,\dots,n,\label{eq:1}
\end{equation}
where $\mathcal{I}$ represents a function support range, $m$ is a smooth and unknown real-valued link function, $\int_{\mathcal{I}}\mathcal{X}_i(t)\beta(t)dt=X_i$ denotes the single index, $\beta(t)$ is an unknown regression coefficient function representing a functional direction that explains the response, and errors $(\varepsilon_1,\varepsilon_2,\dots,\varepsilon_n)$ are possibly correlated errors with unknown error density, denoted by $f(\varepsilon)$, and $\text{E}(\varepsilon_i|X_i) = 0$. 

In equation~\eqref{eq:1}, $m(\cdot)$ can capture a possible non-linear relationship between the single index and response, and it can be the conditional mean \citep{FV06}, conditional quantile \citep{FRV05}, or conditional mode \citep{FLV05}. In this paper, we consider the conditional mean, as it is widely studied in the nonparametric functional data analysis literature. The conditional mean can be estimated via the univariate Nadaraya-Watson (NW) estimator, given by
\begin{equation*}
\widehat{m}(x) = \frac{\sum^n_{i=1}K\left(\frac{X_i - x}{h}\right)y_i}{\sum^n_{i=1}K\left(\frac{X_i - x}{h}\right)},
\end{equation*}
where $K(\cdot)$ is a symmetric kernel function, such as the Gaussian kernel function considered in this paper. Since the mean and error terms are real-valued variables, we use the Gaussian kernel function for both the regression function and error density. As pointed out by \citet[][Chapter 2]{FG96}, the choice of kernel function is not so important in comparison to the bandwidth parameter $h\in R$. The bandwidth parameter often determines the estimation accuracy of the NW kernel estimator.

With the estimated regression functions, the residuals are then used as a proxy for possibly correlated errors \citep[see also][]{Efromovich05}, given by
\begin{align}
\widehat{\varepsilon}_i &= y_i - \widehat{m}(X_i) \notag\\
&= y_i - \widehat{m}\left[\int_{\mathcal{I}} \mathcal{X}_i(t)\widehat{\beta}(t)dt\right], \qquad  i=1,\dots,n. \label{eq:err_ind}
\end{align}

The performance of the functional nonparametric regression crucially depends on the accurate estimation of bandwidth parameter $h$. While \cite{BFR+07} and \cite{RV07} considered a functional version of the cross-validation method for selecting the optimal bandwidth, \cite{Shang13} proposed a Bayesian bandwidth estimation method. In this paper, we extend the model from nonparametric regression to a single index model. Also, we extend the likelihood in the Bayesian bandwidth estimation method from iid error to correlated error structure. 

\subsection{Functional single index model}\label{sec:2.2}

In equation~\eqref{eq:1}, the explanatory functional data are estimated as curves using basis function approximation; that is $\mathcal{X}(t) = \sum^{\infty}_{k=1}c_k\phi_k(t)$ with basis function $\phi_k(t)$ and their associated basis function coefficient $c_k$. Similarly, $\beta(t)$ can be decomposed into $\beta(t) = \sum^{\infty}_{w=1}s_w\psi_w(t)$. The choice of basis function (e.g., $B$-splines, Fourier series, wavelets, principal components) is based on the features of the functional data, such as the periodicity of the data. Because of orthonormality, we consider functional principal components. Thus, equation~\eqref{eq:1} can be expressed as
\begin{align*}
y_i &= m\left[\int \sum^{\infty}_{w=1}\sum^{\infty}_{k=1}c_k s_w\phi_k(t)\psi_w(t)dt\right] + \varepsilon_i \\
&= m\left[\sum^{\infty}_{w=1}\sum^{\infty}_{k=1}c_ks_w \int \phi_k(t)\psi_w(t)dt\right] + \varepsilon_i  \\
&= m\left[\sum^{\infty}_{w=1}\sum^{\infty}_{k=1}c_ks_w\right] + \varepsilon_i. 
\end{align*}

While the errors are treated as iid in~\eqref{eq:err_ind}, we model possible time-series dependence among $(\varepsilon_1,\dots,\varepsilon_n)$. We consider a stationary autoregressive model of order $p$, given by
\begin{equation*}
\varepsilon_j = \sum_{\omega=1}^p\rho_{\omega} \varepsilon_{j-\omega} + \eta_j, \qquad j=p+1,\dots,n,
\end{equation*}
where $\bm{\rho}=(\rho_1, \rho_2,\dots, \rho_p)^{\top}$ denotes a vector of the autoregression parameters, and $\eta_j$ denotes iid errors with zero mean and variance $\sigma_{\epsilon}^2$. The order of $p$ can be determined via the Akaike information criterion corrected for a finite sample size \citep[e.g.,][]{HT89}. Note that the AR order affects the error term, thus it must be pre-determined before constructing the posterior density. In many empirical studies, we find the AR(1) is sufficient to capture the temporal dependence in the error term, and we observe
\begin{align*}
\text{E}\Big[\varepsilon_j^2\Big] = \frac{\sigma_{\epsilon}^2}{1-\rho^2} \\
\text{E}\Big[\varepsilon_j\varepsilon_k\Big] = \frac{\rho^{|j-k|}\sigma_{\epsilon}^2}{1-\rho^2}.
\end{align*}
More generally, let $\bm{\varepsilon} = (\varepsilon_2, \dots, \varepsilon_n)$ and we have
\begin{equation*}
\text{E}\left[\bm{\varepsilon}\bm{\varepsilon}^{\top}\right] = \frac{\sigma_{\epsilon}^2}{1-\rho^2}\left( \begin{array}{ccccc}
1 & \rho & \rho^2 & \cdots & \rho^{n-1} \\
\rho & 1 & \rho & \cdots & \rho^{n-2} \\
\vdots & \vdots & \vdots & \cdots & \vdots \\
\rho^{n-1} & \rho^{n-2} & \rho^{n-3} & \cdots & 1 \end{array} \right). 
\end{equation*}
To avoid singularity, $\rho$ can not be $\pm 1$. For higher orders of AR process, a Yule-Walker equation may be used to estimate higher orders of autocorrelation parameters.

\subsection{Bayesian bandwidth estimation}\label{sec:2.3}

Following the early work by \cite{Shang13}, the Bayesian bandwidth estimation method starts with error density. The unknown error density $f(\varepsilon)$ can be approximated by a location-mixture Gaussian density, given by
\begin{align*}
  f(\eta;b) &= \frac{1}{n-p}\sum^n_{j=p+1}\frac{1}{b}\phi\left(\frac{\eta-\eta_j}{b}\right), \\
  f(\eta;b,\rho)&= \frac{1}{n-p}\sum^n_{j=p+1}\frac{1}{b}\phi\left[\frac{\eta-\left(\varepsilon_j - \sum^p_{\omega=1}\rho_{\omega} \varepsilon_{j-\omega}\right)}{b}\right],
\end{align*}
where $\phi(\cdot)$ is the probability density function of the standard Gaussian distribution, and the Gaussian densities have means at $\eta_j$ and a common standard deviation $b$.

Since the errors are unknown in practice, we approximate them by residuals obtained from the functional NW estimator of conditional mean. Given bandwidths $h$ and $b$ and autoregression parameter $\bm{\rho}$, the kernel likelihood of $\bm{y}$ is given by
\begin{equation*}
  \widehat{L}\left(\bm{y}|h, b, \bm{\rho}\right) = \prod^n_{i=p+1}\Bigg\{\frac{1}{n-p-1}\sum^n_{\substack{j=p+1\\ j\neq i}}\frac{1}{b}\phi\left[\frac{\left(\widehat{\varepsilon}_{i}-\sum^p_{\omega=1}\rho_{\omega}\widehat{\varepsilon}_{i-\omega}\right)-\left(\widehat{\varepsilon}_{j}-\sum^p_{\omega=1}\rho_{\omega}\widehat{\varepsilon}_{j-\omega}\right)}{b}\right]\Bigg\}.
\end{equation*}

\subsubsection{Prior density}

We discuss the choice of prior density for the bandwidths. Let $\pi(h^2)$  and $\pi(b^2)$ be the prior densities of the squared bandwidths $h$ and $b$. Since $h^2$ and $b^2$ can be considered as a variance parameter in the Gaussian distribution, we consider the conjugate prior of $h^2$ and $b^2$. Let the prior densities of $h^2$ and $b^2$ be inverse Gamma densities, denoted as $\text{IG}(\alpha_h, \beta_h)$ and $\text{IG}(\alpha_b, \beta_b)$, respectively. Since $-1<\rho<1$ is a correlation parameter, we consider a uniform density as prior density. The prior densities of the bandwidth parameters and autocorrelation parameters can be expressed as
\begin{align*}
\pi(h^2) &= \frac{(\beta_h)^{\alpha_h}}{\Gamma(\alpha_h)}\left(\frac{1}{h^2}\right)^{\alpha_h+1}\exp\left(-\frac{\beta_h}{h^2}\right), \\
\pi(b^2) &= \frac{(\beta_b)^{\alpha_b}}{\Gamma(\alpha_b)}\left(\frac{1}{b^2}\right)^{\alpha_b+1}\exp\left(-\frac{\beta_b}{b^2}\right), \\
\pi(\bm{\rho}) & = \left(\frac{1}{2}\right)^p
\end{align*}
where $\alpha_h = \alpha_b = 1$ and $\beta_h = \beta_b = 0.05$ as hyper-parameters \citep[see also][]{Geweke10}. In Section~\ref{sec:3.4}, we also consider the Cauchy prior density for bandwidth parameters $h$ and $b$.

\subsubsection{Posterior density}

According to Bayes theorem, the posterior of $h^2_n$, $b^2_n$ and $\rho$ is approximated by (up to a normalizing constant)
\begin{equation}
\pi\left(h_n^2, b_n^2, \bm{\rho}\big|\bm{y}\right) \propto \widehat{L}\Big(\bm{y}\big|h_n^2, b_n^2, \bm{\rho}\Big)\pi\Big(h^2\Big)\pi\Big(b^2\Big)\pi\Big(\bm{\rho}\Big), \label{eq:bayes}
\end{equation}
where $\widehat{L}\left(\bm{y}|h_n^2, b_n^2, \bm{\rho}\right)$ is the approximate kernel likelihood function with squared bandwidths. The parameters are sampled from its posterior density and estimated by the means of Markov chain Monte Carlo (MCMC). In essence, the Monte Carlo method sets up a Markov chain so that its stationary distribution is the same as the posterior density. When the Markov chain converges, the ergodic averages of the simulated realizations are treated as the estimated parameter values. For a detailed exposition of the use of MCMC method, refer to the seminal works by \cite{Geweke99}, \cite{GRS96} and \cite{RC10}.

\subsubsection{Adaptive random-walk Metropolis algorithm}

From equation~\eqref{eq:bayes}, we use a generic algorithm known as the adaptive random-walk Metropolis algorithm of \cite{GFS16} to sample $\left(h_n^2, b_n^2, \bm{\rho}\right)$ jointly. The sampling algorithm is described below.
\begin{enumerate}
\item[1)] Specify a Gaussian proposal density, with an arbitrary starting point $b^2_{n,(0)}$, $h^2_{n, (0)}$ and $\bm{\rho}_{n,(0)}$. The starting points can be drawn from a uniform distribution $U(0, 1)$.
\item[2)] At the $k$\textsuperscript{th} iteration, the current state $b^2_{n,(k)}$ is updated as $b^2_{n, (k)} = b^2_{n, (k-1)} + \tau_{(k-1)}\vartheta$, where $\vartheta \sim N(0,1)$, and $\tau_{(k-1)}$ is an adaptive tuning parameter with an arbitrary initial value $\tau_{(0)}$.
\item[3)] The updated $b^2_{n, (k)}$ is accepted with probability
\begin{equation*}
\min \left\{\frac{\pi\left(b^2_{n, (k)}, h^2_{n, (k-1)}, \bm{\rho}_{n,(k-1)}\big|\bm{y}\right)}{\pi\left(b^2_{n, (k-1)}, h^2_{n, (k-1)}, \bm{\rho}_{n, (k-1)}\big|\bm{y}\right)}, 1\right\},
\end{equation*}
where $\pi$ represents the posterior density.
\item[4)] By using the stochastic search algorithm of \cite{RM51}, the adaptive tuning parameter is
\[ \tau_{(k)} = \left\{ \begin{array}{ll}
         \tau_{(k-1)}+c(1-p)/k & \mbox{if $b^2_{n,(k)}$ is accepted};\\
        \tau_{(k-1)}-cp/k & \mbox{if $b^2_{n,(k)}$ is rejected},\end{array} \right. \] 
where $c = \frac{\tau_{(k-1)}}{p(1-p)}$ is a varying constant, and $p=0.44$ is the optimal acceptance probability for drawing one parameter \citep{RR09}.
\item[5)] Repeat Steps 1)-4) for $h^2_{n, (k)}$, conditional on $b^2_{n, (k)}, \bm{\rho}_{n, (k)}$ and $\bm{y}$. Similarly, repeat steps 1)-4) for $\bm{\rho}_{n, (k)}$, conditional on $h_{n,(k)}^2, b_{n,(k)}^2$ and $\bm{y}$.
\item[6)] Repeat Steps 1)-5) for $M+N$ times, discard $\left(h^2_{n, (0)}, b^2_{n, (0)}, \bm{\rho}_{n, (0)}\right)$, $\left(h^2_{n, (1)}, b^2_{n, (1)}, \bm{\rho}_{n, (1)}\right)$, $\dots$, \linebreak $\left(h^2_{n, (M)}, b^2_{n, (M)}, \bm{\rho}_{n, (M)}\right)$ for burn-in in order to let the effects of the transients wear off, estimate $\widehat{h}_n^2 = \frac{\sum^{M+N}_{k=M+1}h^2_{n, (k)}}{N}$, $\widehat{b}^2_n = \frac{\sum^{M+N}_{k=M+1}b^2_{n, (k)}}{N}$ and $\widehat{\bm{\rho}}_{n} = \frac{\sum^{M+N}_{k=M+1}\bm{\rho}_{n,(k)}}{N}$. The burn-in period is taken to be $M=1,000$ iterations, and the number of iterations after the burn-in period is $N=10,000$ iterations. The analytical form of the kernel-form error density can be derived from $\widehat{h}_n^2$, $\widehat{b}_n^2$ and $\widehat{\bm{\rho}}_n$.         
\end{enumerate}

The mixing performance of the sample paths can be measured by total standard error (SE), and batch mean SE, from which we can also calculate the simulation inefficiency factor \citep[see also][]{KSC98, MY00}. The simulation inefficiency factor can be interpreted as the number of draws needed to have iid observations. 

\section{Simulation study}\label{sec:4}

The main goal of this section is to illustrate the proposed methodology through simulated data. One way to do this consists of comparing the true regression function with the estimated regression function and comparing the true error density with the estimated error density.

\subsection{Criteria for assessing estimation accuracy}

To measure the estimation accuracy between $m(\mathcal{X})$ and $\widehat{m}(\mathcal{X})$, we first approximate the mean squared error (MSE) given by
\begin{align*}
\text{MSE} &= \text{E}[m(\mathcal{X}) - \widehat{m}(\mathcal{X})]^2.
\end{align*}
Averaged across 100 replications, the averaged MSE (AMSE) is used to assess the estimation accuracy of regression function. This is defined as
\begin{align*}
\text{AMSE} &= \frac{1}{B}\sum^B_{b=1}\text{MSE}_b,
\end{align*}
where $B=100$ represents the number of replications.

To measure the difference between $f(\varepsilon)$ and $\widehat{f}(\varepsilon)$, we first approximate the mean integrated squared error (MISE). This is given by
\begin{align*}
\text{MISE}\Big[\widehat{f}(\varepsilon)\Big] &= \int^b_a \Big[f(\varepsilon) - \widehat{f}(\varepsilon)\Big]d\varepsilon,
\end{align*}
for $\varepsilon \in [a,b]$. For each replication, the MISE can be approximated at 1,001 grid points bounded between an interval, such as $[-5,5]$. These can be expressed as
\begin{align*}
\text{MISE}\Big[\widehat{f}(\varepsilon)\Big] &\approx \frac{1}{100}\sum^{1001}_{i=1}\left\{f\Big[-5+\frac{(i-1)}{100}\Big]-\widehat{f}\Big[-5+\frac{(i-1)}{100}\Big]\right\}^2.
\end{align*} 
Averaged across 100 replications, the averaged MISE (AMISE) is used to assess the estimation accuracy of error density. This is defined as
\begin{equation*}
\text{AMISE} =\frac{1}{B}\sum^B_{b=1}\text{MISE}_b.
\end{equation*}

\subsection{Smooth curves}\label{sec:smooth_curves}

\begin{figure}[!ht]
  \centering
  \subfloat[$n=60$]
{\includegraphics[width=8.5cm]{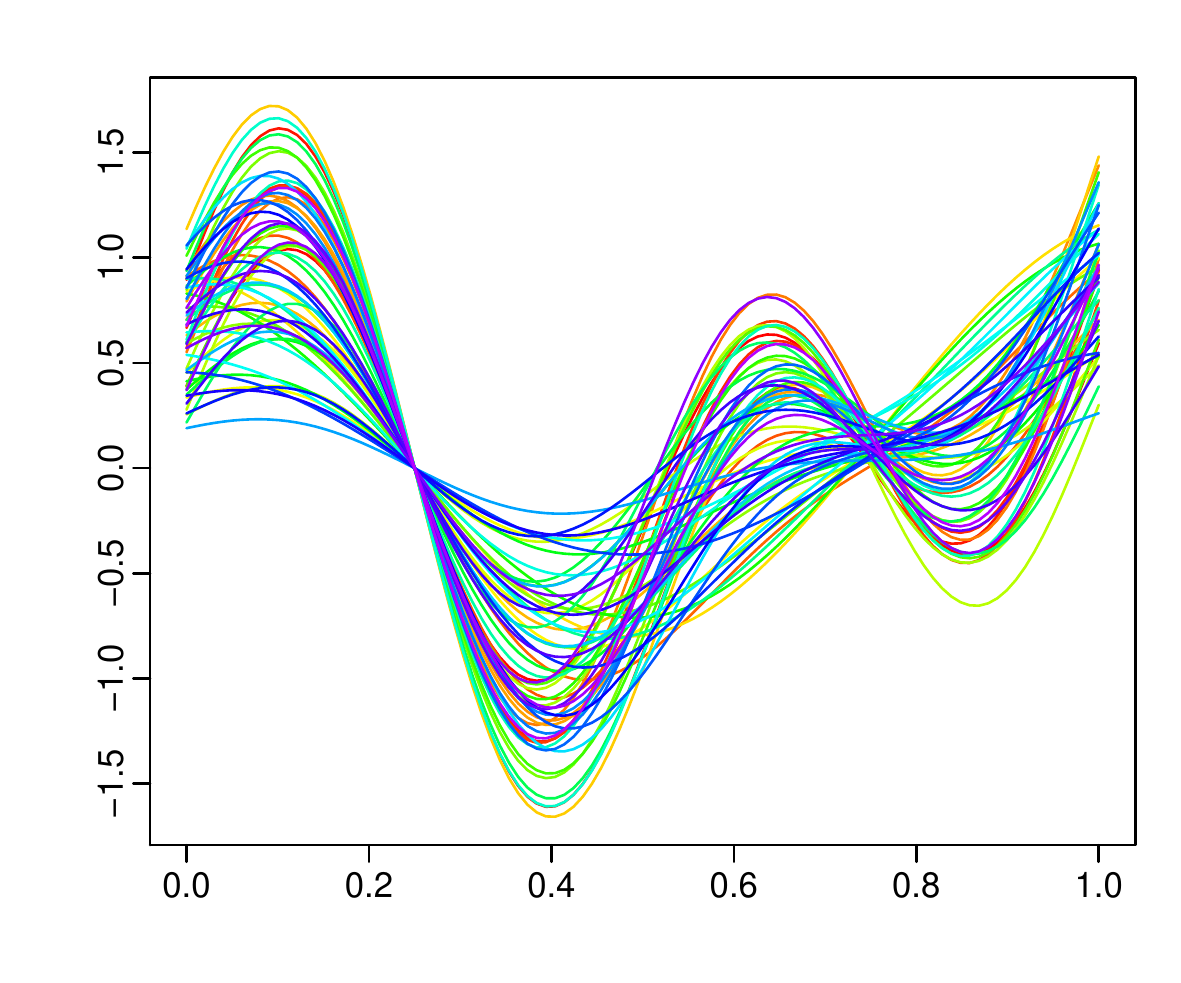}}
\qquad
\subfloat[$n=120$]
{\includegraphics[width=8.5cm]{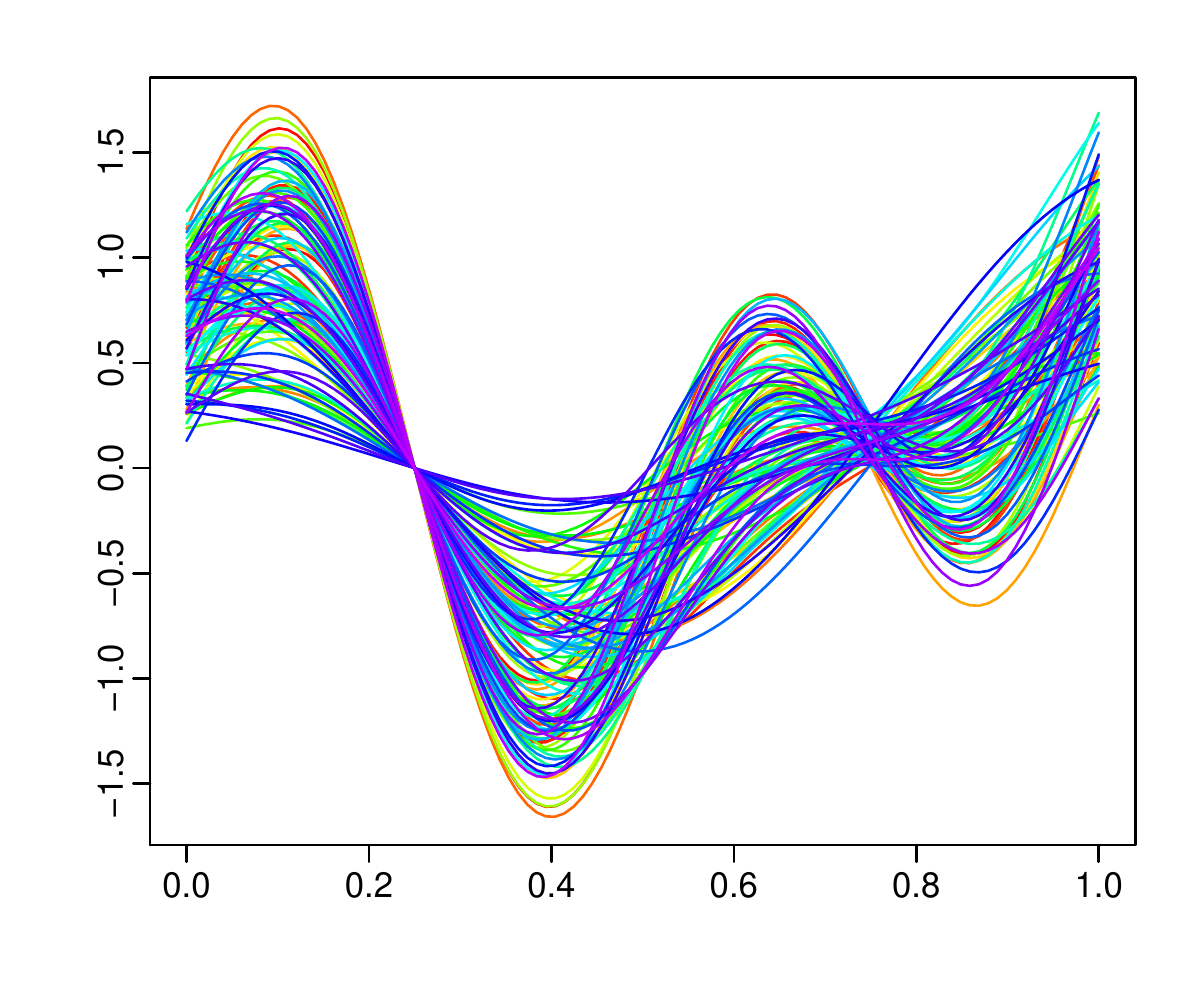}}
  \caption{\small Simulated smooth curves. Based on a rainbow color palette, the curves from the distant past are shown in red, while the most recent curves are shown in purple \citep[c.f.,][]{HS10}.}\label{fig:1}
\end{figure}

First of all, we build a sample of $n$ curves as follows
\begin{equation}
\mathcal{X}_i(t_j) = a_i\cos(2\pi t_j)+b_i\sin(4\pi t_j)+2c_i(t_j-0.25)(t_j-0.5), \qquad i=1,2,\dots,n, \label{eq:smooth_curves}
\end{equation}
where $t$ represents the function support range and $0\leq t_1\leq t_2\dots\leq t_{100}\leq 1$ are equispaced points within the function support range, while $a_i, b_i, c_i$ are independently drawn from a uniform distribution on $[0,1]$. Figure~\ref{fig:1} presents two rainbow plots of the simulated smooth curves for one replication with $n=60$ and $n=120$, respectively.

Once the curves are defined, we simulate a functional single index model to compute the response variable in the following steps:
\begin{itemize}
\item choose one regression coefficient function $\beta(\cdot)$ and let it be $\beta(\cdot) = \sin(\pi t)$;
\item compute the inner products $X_i = \langle \beta, \mathcal{X}_i\rangle$ for $i=1,\dots,n$;
\item choose one link function $m(\cdot)$ and let it be $m(\cdot)=100\times (X-0.15)^3$;
\item generate $\varepsilon_1,\varepsilon_2,\dots,\varepsilon_n$ which are either independently drawn from normal distribution with three different choices of signal-to-noise ratio (Tables~\ref{tab:1} and~\ref{tab:2}) or follows AR$_{0.8}(1)$ (Tables~\ref{tab:11} and~\ref{tab:22}). The signal-to-noise ratio is defined as $\sigma^2_{\text{signal}}/\sigma^2_{\text{noise}}$. Denote $\xi$ as the inverse signal-to-noise ratio. In order to highlight possible non-normality of the error density, we consider different mixture normal distributions previously introduced by \cite{MW92} and the results are reported in the supplement;
\item compute the corresponding responses: $y_i = m\left(\int_{\mathcal{I}}\mathcal{X}_i(t)\beta(t)dt\right) + \varepsilon_i$, for $i=1,2,\dots, n$.
\end{itemize}

\paragraph{Estimating the regression function} For a given curve $\mathcal{X}$ and an estimated bandwidth $h$, we compute the in-sample estimation discrepancy between $m(\mathcal{X})$ and $\widehat{m}(\mathcal{X})$. To do that, we use the following Monte Carlo scheme:
\begin{itemize}
  \item build 100 replications $\{(\mathcal{X}_i^s, y_i^s)_{i=1,\dots,n_1}\}_{s=1,\dots,100}$, where $s$ denotes one sample replication; and $n_1$ denotes the number of training samples;
  \item compute 100 estimates $\text{MSE}_s = \frac{1}{n_1}\sum_{i=1}^{n_1}[m(\mathcal{X}_i)-\widehat{m}_h^s(\mathcal{X}_i)]^2_{s=1,\dots,100}$, where $\widehat{m}_h^s(\mathcal{X}_i)$ is the functional NW estimator of the regression function for the $i$\textsuperscript{th} sample curve computed over the $s^{\text{th}}$ replication;
  \item obtain the AMSE by averaging across 100 replications of MSE.
\end{itemize}

\paragraph{Predicting the regression function} For a given new curve $\mathcal{X}_{\text{new}}$ and an estimated bandwidth $h$, we compute the out-of-sample prediction discrepancy between $m(\mathcal{X}_{\text{new}})$ and $\widehat{m}(\mathcal{X}_{\text{new}})$. To do that, we use the following Monte Carlo scheme:
\begin{itemize}
\item build 100 replications $\{(\mathcal{X}_i^s, y_i^s)_{i=1,\dots,n_1}\}_{s=1,\dots,100}$;
\item compute 100 estimates of the mean square prediction error $\text{MSPE}_s = \frac{1}{n_2}\sum_{i=1}^{n_2}[m(\mathcal{X}_{\text{new},i})-\widehat{m}_h^s(\mathcal{X}_{\text{new}, i})]^2_{s=1,\dots,100}$, where $n_2$ denotes the number of testing samples, and $\widehat{m}_h^s(\mathcal{X}_{\text{new},i})$ is the functional NW estimator of the regression function for the $i$\textsuperscript{th} sample curve computed over the $s^{\text{th}}$ replication;
\item obtain the averaged MSPE (AMSPE) by averaging across 100 replications of MSPE.
\end{itemize}

Table~\ref{tab:1} compares the AMSE, AMISE, and AMSPE between the functional single index model and the nonparametric functional regression model, under an iid Gaussian error density. In the nonparametric functional regression, we study a range of semi-metrics, such as the semi-metric based on 1\textsuperscript{st} and 2\textsuperscript{nd} derivatives and the semi-metric based on functional principal component analysis with different numbers of retained components \citep[][Chapter 3]{FV06}. As the signal-to-noise ratio decreases (i.e., $\xi$ value increases), the regression function becomes harder to estimate, thus the AMSE and AMSPE increase. Because the regression function becomes harder to estimate, the resultant residuals vary greatly, and this helps the estimation of error density. Thus the AMISE decreases. The functional single index model produces much more accurate estimation and prediction accuracies than any nonparametric functional regression.

\begin{center}
\tabcolsep 0.095in
\renewcommand{\arraystretch}{0.61}
\begin{longtable}{@{}lllccccccc@{}}
  \caption{\small Estimation accuracy of the regression function between the functional single index model and the nonparametric functional regression with different choices of semi-metrics, for a set of smoothed curves under the Gaussian error density with the iid error structure and three different signal-to-noise ratios $\xi$. The numbers in parenthesis represent the sample standard deviation of the squared errors. The text in bold represents the minimal AMSE, AMISE, and AMSPE. NFR denotes the nonparametric functional regression, while FSIM denotes the functional single index model.}\label{tab:1}\\
  \toprule
  & & & \multicolumn{3}{c}{$n=60$}  & & \multicolumn{3}{c}{$n=120$} \\
  Error & Model & Semi-metric& $\xi=0.1$ & $\xi=0.5$ & $\xi=0.9$ & & $\xi=0.1$ & $\xi=0.5$ & $\xi=0.9$ \\\toprule
\endfirsthead
  & & & \multicolumn{3}{c}{$n=60$}  & & \multicolumn{3}{c}{$n=120$} \\
  Error & Model & Semi-metric& $\xi=0.1$ & $\xi=0.5$ & $\xi=0.9$ & & $\xi=0.1$ & $\xi=0.5$ & $\xi=0.9$ \\\toprule
\endhead
\hline \multicolumn{10}{r}{{Continued on next page}} \\
\endfoot
\endlastfoot
AMSE  & NFR & $\text{deriv}_1$ & 0.4501 & 0.5785 & 0.6699 & & 0.3693 & 0.4476 & 0.5151  \\
  & &                  & (0.1600) & (0.1968) & (0.2237) & & (0.0821) & (0.1043) & (0.1203) \\\\
 &  & $\text{deriv}_2$ &   0.9746 &  1.1374 &  1.3146 & & 0.5358 & 0.6657 & 0.7637  \\
&     &              & (0.3197) & (0.3114) & (0.4453) & & (0.2335) & (0.2122) &  (0.1917) \\\\
& &   $\text{pca}_1$ & 0.6405 & 0.6983 & 0.7566 & & 0.5974 &  0.6263 &  0.6532 \\
&    &               & (0.1430) & (0.1576) & (0.1836) & & (0.1063) & (0.1072) &  (0.1135) \\\\
& &   $\text{pca}_2$ & 0.3911 & 0.4583 & 0.5248 &  & 0.3551 & 0.3900 &  0.4232 \\
&    &               & (0.0786) & (0.0988) & (0.1514) & & (0.0603) & (0.0664) & (0.0783) \\\\
& &   $\text{pca}_3$ & 0.2042 & 0.3071 & 0.3774 & & 0.1274 & 0.2075 & 0.2607 \\
&    &               & (0.0470) & (0.0764) & (0.1047) & & (0.0224) & (0.0407) &  (0.0566) \\\\
&   FSIM    &  & \textBF{0.0249} & \textBF{0.0893} & \textBF{0.1484}  & & \textBF{0.0149}  & \textBF{0.0518} & \textBF{0.0840} \\
&   &                  & (0.0110) & (0.0455) & (0.0853) & & (0.0067) & (0.0254) & (0.0423)  \\\\
AMISE & NFR & deriv$_1$ & 0.2811 & 0.0446 & 0.0226 & & 0.2317 & 0.0328 & 0.0163 \\
		&		&	& (0.0905) & (0.0214) & (0.0143) & & (0.0582) & (0.0143) & (0.0090)		\\\\
		& & deriv$_2$ & 0.4478 & 0.0870 & 0.0443 & & 0.2871 & 0.0486 & 0.0246 \\
		&		&	& (0.1402) & (0.0314) & (0.0232) & & (0.0992) & (0.0213) & (0.0113)		\\\\
		& & pca$_1$ & 0.3464 & 0.0593 & 0.0293 & & 0.3197 & 0.0473 & 0.0221  \\
		&		&	& (0.0817) & (0.0224) & (0.0132) & & (0.0624) & (0.0187) & (0.0112)	\\\\
		& & pca$_2$ & 0.2662 & 0.0421 & 0.0236 & & 0.2353 & 0.0308 & 0.0154  \\
		&		&	& (0.0857) & (0.0212) & (0.0130) & & (0.0512) & (0.0141) & (0.0088)	\\\\
		& & pca$_3$ & 0.1647 & 0.0322 & 0.0199 & & 0.0980 & 0.0192 & 0.0116 \\
		&		&	& (0.0735) & (0.0183) & (0.0121) & & (0.0358) & (0.0127) & (0.0088)	\\\\
		& FSIM 	&	& \textBF{0.0391} & \textBF{0.0206} & \textBF{0.0150} & & \textBF{0.0226} & \textBF{0.0105} & \textBF{0.0082}	\\
		&		&	& (0.0370) & (0.0235) & (0.0150) & & (0.0280) & (0.0113) & (0.0106)	\\\\		
AMSPE & NFR & deriv$_1$ & 0.4801 & 0.5587 & 0.6863 & & 0.4229 &  0.4953 &  0.5735 \\
		&		&  & (0.2721) &	(0.3285) & (0.4800) & & (0.1919) & (0.2006) & (0.2312) \\\\
		& & deriv$_2$ & 0.9635 & 1.1042 & 1.2938 & & 0.5911 & 0.6988 & 0.7654  \\
		&		& & (0.4800) & 	(0.5894) & (0.6727) & & (0.3250) & (0.3940) & (0.2709) \\\\
		& & pca$_1$ & 0.6181 & 0.6689 & 0.7210 & & 0.6189 & 0.6424 & 0.6720 \\
		&		&  & (0.3073) & (0.3495) & (0.3792) & & (0.2137) & (0.2076) &	(0.2116)	\\\\
		& & pca$_2$ & 0.3998 &  0.4622 & 0.5332 & & 0.3749 & 0.4113 & 0.4466 \\
		&		& 	& (0.2298) & (0.2845) & (0.3460) & & (0.1508) & (0.1587) & (0.1667)	\\\\
		& & pca$_3$ & 0.2027 & 0.3014 & 0.3679 & & 0.1296 & 0.2134 & 0.2647 \\
		&		& 	& (0.1391) & (0.1989) & (0.2486) & & (0.0702) & (0.1072) & (0.1154) 	\\\\	
		& FSIM & & \textBF{0.0317} & \textBF{0.0961} & \textBF{0.1574} & & \textBF{0.0177} & \textBF{0.0530} & \textBF{0.0833} \\
		&		&  & (0.0278) & (0.0696) & (0.1199) & & (0.0124) & (0.0303) &	(0.0476)	\\\bottomrule
\end{longtable}
\end{center}

When errors are simulated from a Gaussian density with the AR$_{\rho=0.8}(1)$ structure, Table~\ref{tab:11} presents the AMSE, AMISE, and AMSPE for the functional single index model and the nonparametric functional regression models. As the signal-to-noise ratio decreases (i.e., $\xi$ value increases), the AMSE and AMSPE increase whereas the AMISE decreases. The functional single index model performs better than the nonparametric functional regression in all criteria. Compared to the results under the iid error structure in Table~\ref{tab:1}, the AMSE, AMISE, and AMSPE increase under the AR(1) error structure.

\begin{center}
\tabcolsep 0.095in
\renewcommand{\arraystretch}{0.61}
\begin{longtable}{@{}lllccccccc@{}}
  \caption{\small Estimation accuracy of the regression function between the functional single index model and the nonparametric functional regression with different choices of semi-metrics, for a set of smoothed curves under the Gaussian error density with the AR$_{\rho=0.8}(1)$ error structure and three different signal-to-noise ratios $\xi$.}\label{tab:11}\\
  \toprule
  & & & \multicolumn{3}{c}{$n=60$}  & & \multicolumn{3}{c}{$n=120$} \\
  Error & Model & Semi-metric& $\xi=0.1$ & $\xi=0.5$ & $\xi=0.9$ & & $\xi=0.1$ & $\xi=0.5$ & $\xi=0.9$ \\\toprule
\endfirsthead
  & & & \multicolumn{3}{c}{$n=60$}  & & \multicolumn{3}{c}{$n=120$} \\
  Error & Model & Semi-metric& $\xi=0.1$ & $\xi=0.5$ & $\xi=0.9$ & & $\xi=0.1$ & $\xi=0.5$ & $\xi=0.9$ \\\toprule
\endhead
\hline \multicolumn{10}{r}{{Continued on next page}} \\
\endfoot
\endlastfoot
AMSE & NFR & deriv$_1$ & 0.5172 & 0.9146 & 1.2727 & & 0.4260 & 0.6740 & 0.9210 \\
	&		&		& (0.1515) & (0.3859) & (0.6770) & & (0.0962) & (0.2412) & (0.4230) \\
\\
	&		& deriv$_2$ & 0.9934 & 1.3515 & 1.7948 & & 0.6421 & 0.9364 & 1.1731 \\
	&		&		&	(0.3540) & (0.4307) & (0.7432) & & (0.2645) & (0.2878) & (0.4009) \\
\\
	&		& pca$_1$ & 0.7052 & 1.0088 & 1.2991 & & 0.6342 & 0.8158 & 0.9883 \\
	&		&		& (0.1605) & (0.3858) & (0.6482) & & (0.1137) & (0.2495) & (0.4229) \\
\\
	&		& pca$_2$ & 0.4562 & 0.7752 & 1.0675 & & 0.3947 & 0.5846 & 0.7573 \\
	&		&		& (0.1177) & (0.3997) & (0.6964) & & (0.0716) & (0.2301) & (0.4049) \\
\\
	&		& pca$_3$ & 0.2951 & 0.6548 & 0.9708 & & 0.1907 & 0.4314 & 0.6242 \\
	&		&		& (0.0971) & (0.3997) & (0.6993) & & (0.0566) & (0.2272) & (0.4029) \\
\\
	&	FSIM	& & \textBF{0.0885} & \textBF{0.3819} & \textBF{0.6612} & & \textBF{0.0541} & \textBF{0.2293} & \textBF{0.3941} \\
	&			&	& (0.0674) & (0.3455) & (0.6185) & & (0.0484) & (0.2355) & (0.4225) \\
\\				
AMISE & NFR & deriv$_1$ & 0.3302 & 0.0964 & 0.0670 & & 0.3033 & 0.0855 & 0.0564 \\
	&		&		& (0.0810) & (0.0459) & (0.0389) & & (0.0644) & (0.0300) & (0.0209) \\
\\	
	&	& deriv$_2$ & 0.4645 & 0.1155 & 0.0724 & & 0.3604 & 0.1010 & 0.0635 \\
	&		&		& (0.1262) & (0.0442) & (0.0310) & & (0.1108) & (0.0345) & (0.0240) \\
\\
	& 	& pca$_1$ & 0.3878 & 0.1053 & 0.0696 & & 0.3658 & 0.0946 & 0.0610  \\
	&	&		& (0.0818) & (0.0385) & (0.0317) & & (0.0611) & (0.0286) & (0.0218) \\
\\	
	& 	& pca$_2$ & 0.3206 & 0.0926 & 0.0641 & & 0.3022 & 0.0841 & 0.0563 \\
	&	&		& (0.0875) & (0.0423) & (0.0329) & & (0.0675) & (0.0309) & (0.0229) \\
\\	
	& 	& pca$_3$ & 0.2579 & 0.0870 & 0.0621 & & 0.2214 & 0.0755 & 0.0534 \\
	&	&		& (0.0918) & (0.0417) & (0.0334) & & (0.0697) & (0.0302) & (0.0222) \\
\\	
	& FSIM & 	& \textBF{0.0549} & \textBF{0.0244} & \textBF{0.0169} & & \textBF{0.0295} & \textBF{0.0106} & \textBF{0.0089} \\
	&		& 	& (0.0545) & (0.0267) & (0.0183) & & (0.0337) & (0.0087) & (0.0083) \\
\\
AMSPE & NFR & deriv$_1$ & 0.5505 & 0.9677 & 1.3164 & & 0.4865 & 0.7456 & 0.9580 \\
		&	&	& (0.3398) & (0.6974) & (0.9196) & & (0.2146) & (0.4049) & (0.5550) \\
\\
	&		& deriv$_2$ & 0.9444 & 1.5785 & 1.7389 & & 0.6885 & 0.9924 & 1.2000 \\
	&		&		& (0.5703) & (1.0303) & (1.1091) & & (0.4270) & (0.4223) & (0.4108) \\
\\
	& 		& pca$_1$ & 0.6884 & 1.0172 & 1.3380 & & 0.6643 & 0.8507 & 1.0240 \\
	&		&		& (0.3966) & (0.7489) & (1.1041) & & (0.2262) & (0.3464) & (0.5025) \\
\\
	& 		& pca$_2$ & 0.4739 & 0.7725 & 1.0610 & & 0.4126 & 0.5957 & 0.7632 \\
	&		&		& (0.3144) & (0.6183) & (0.9628) & & (0.1721) & (0.3147) & (0.4914) \\
\\
	& 		& pca$_3$ & 0.3013 & 0.6442 & 0.9573 & & 0.2004 & 0.4470 & 0.6453 \\
	&		&		& (0.1859) & (0.5323) & (0.8697) & & (0.1006) & (0.2803) & (0.4638) \\
\\
	& FSIM 	& & \textBF{0.0992} & \textBF{0.3900} & \textBF{0.6708} & & \textBF{0.0581} & \textBF{0.2394} & \textBF{0.4104} \\
	&		& & (0.0932) & (0.3898) & (0.6944) & & (0.0503) & (0.2529) & (0.4617) \\
\bottomrule					
\end{longtable}
\end{center}

\subsection{Rough curves}

\begin{figure}[!htbp]
  \centering
  \subfloat[$n=60$]
  {\includegraphics[width=8.55cm]{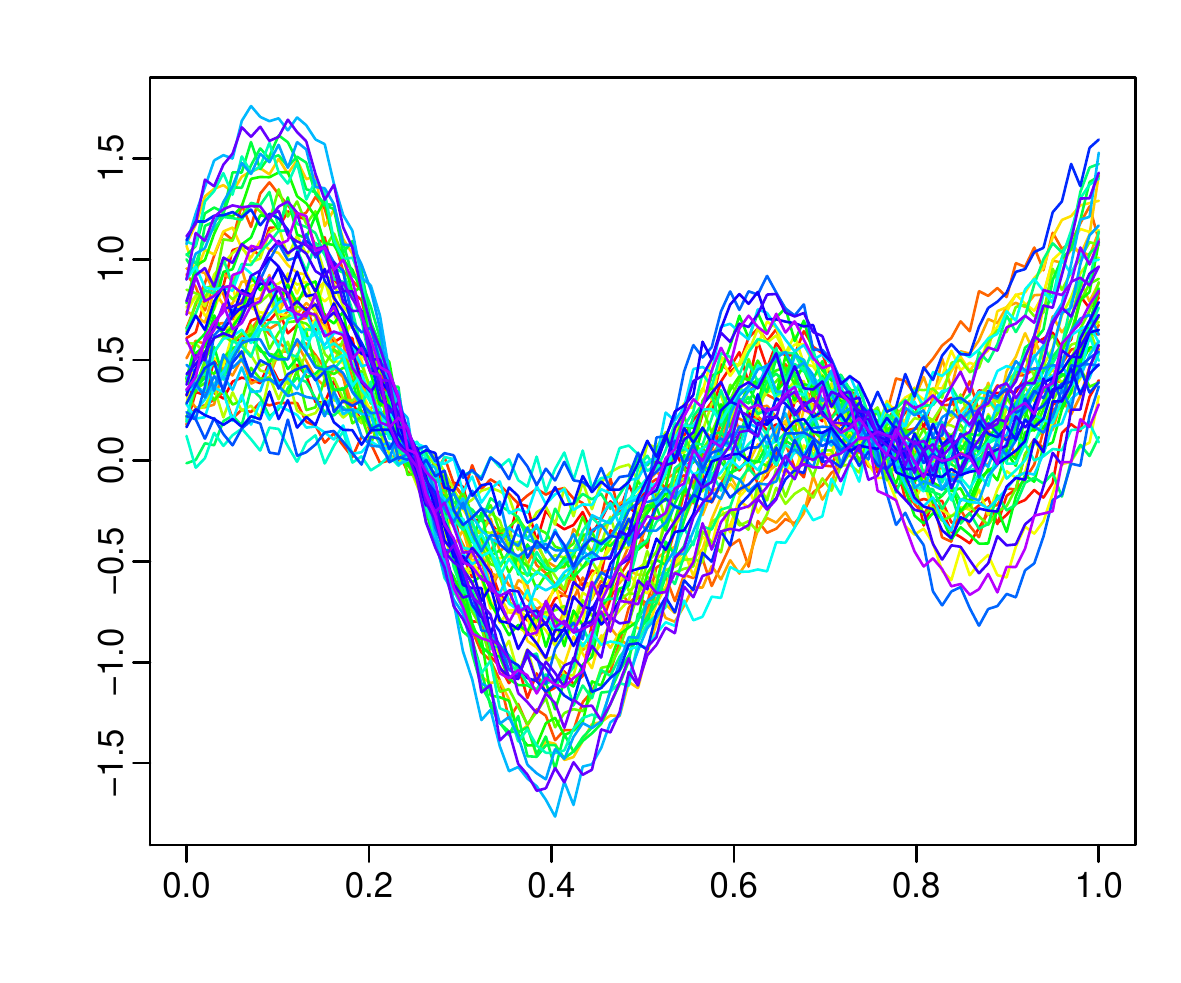}}
  \qquad
  \subfloat[$n=120$]
  {\includegraphics[width=8.55cm]{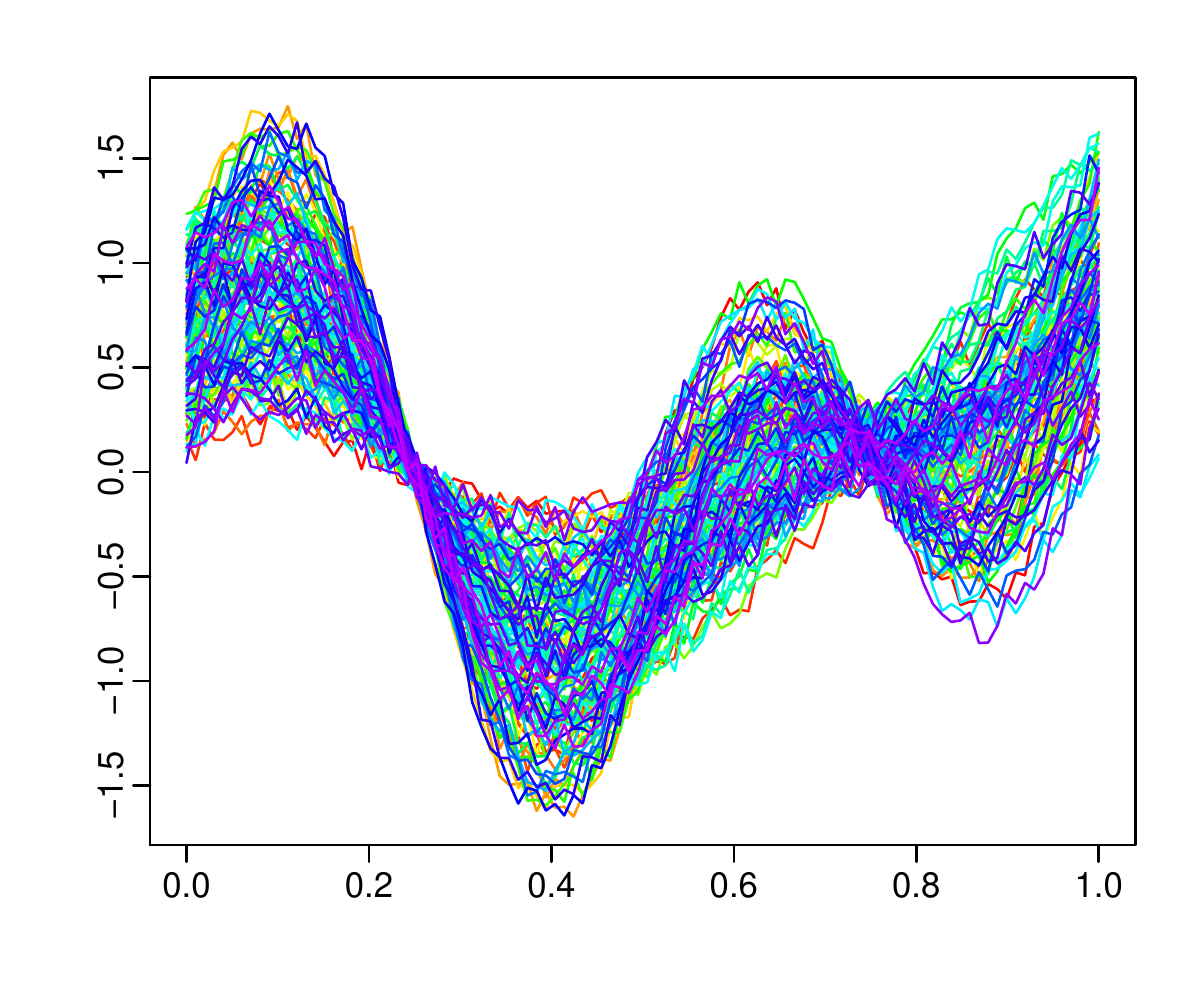}}
  \caption{\small Simulated rough curves.}\label{fig:3}
\end{figure}

In this simulation study, we consider the same functional form as given in equation~\eqref{eq:smooth_curves}, but add one extra variable $d_j\sim U(-0.1, 0.1)$ in the construction of functional curves. This data-generating process has been considered in \cite{Shang14b}. Figure~\ref{fig:3} presents the simulated rough curves for one replication with $n=60$ and $n=120$, respectively. 

Using the same setup as in Section~\ref{sec:smooth_curves}, Table~\ref{tab:2} presents the AMSE, AMISE, and AMSPE for the functional single index model and the nonparametric functional regression models, under an iid Gaussian error density. In the nonparametric functional regression, we study a range of semi-metrics, such as the semi-metric based on 1\textsuperscript{th} and 2\textsuperscript{nd} derivatives and the semi-metric based on the functional principal components. As the signal-to-noise ratio decreases (i.e., $\xi$ value increases), the AMSE and AMSPE increase, whereas the AMISE decreases. The functional single index model produces much more accurate estimation and prediction accuracies than any nonparametric functional regression in all criteria. 

\begin{center}
\tabcolsep 0.095in
\renewcommand{\arraystretch}{0.82}
\begin{longtable}{@{}lllccccccc@{}}
  \caption{\small Estimation accuracy of the regression function between the functional single index model and the nonparametric functional regression with different choices of semi-metrics, for a set of rough curves under the Gaussian error density with the iid error structure and three different signal-to-noise ratios $\xi$.}\label{tab:2}\\
  \toprule
  & & & \multicolumn{3}{c}{$n=60$}  & & \multicolumn{3}{c}{$n=120$} \\
  Error & Model & Semi-metric& $\xi=0.1$ & $\xi=0.5$ & $\xi=0.9$ & & $\xi=0.1$ & $\xi=0.5$ & $\xi=0.9$ \\\toprule
\endfirsthead
  & & & \multicolumn{3}{c}{$n=60$}  & & \multicolumn{3}{c}{$n=120$} \\
  Error & Model & Semi-metric& $\xi=0.1$ & $\xi=0.5$ & $\xi=0.9$ & & $\xi=0.1$ & $\xi=0.5$ & $\xi=0.9$ \\\toprule
\endhead
\hline \multicolumn{10}{r}{{Continued on next page}} \\
\endfoot
\endlastfoot
AMSE & NFR & $\text{deriv}_1$ & 0.5086 & 0.5786 & 0.6532 & & 0.4600 & 0.5095 & 0.5656  \\
& &                  & (0.1290) & (0.1505) & (0.1691) & & (0.0880) & (0.1006) & (0.1147) \\\\
&  & $\text{deriv}_2$ &  1.1826 & 1.1691 & 1.2106 & & 1.0540 & 1.0737 & 1.1051 \\
&    &              & (0.2447) & (0.2496) & (0.2660) & & (0.1191) & (0.1337) & (0.1472)  \\\\
& &   $\text{pca}_1$ & 0.6204 & 0.6720 & 0.7256 &  & 0.6135 & 0.6391 & 0.6632  \\
&    &               & (0.1327) & (0.1353) & (0.1511) & & (0.0926) & (0.0975) & (0.0987)  \\\\
& &   $\text{pca}_2$ & 0.4081 & 0.4800 & 0.5521 & & 0.3667 & 0.4098 & 0.4475 \\
&    &               & (0.0907) & (0.1342) & (0.1702) & & (0.0577) & (0.0649) & (0.0824) \\\\
& &   $\text{pca}_3$ & 0.2050 & 0.3243 & 0.3992 & & 0.1343 & 0.2191 & 0.2736 \\
&    &               & (0.0465) & (0.0900) & (0.1140) & & (0.0231) & (0.0419) & (0.0570) \\\\
&   FSIM    &  & \textBF{0.0297} & \textBF{0.1008} & \textBF{0.1660} & & \textBF{0.0177} & \textBF{0.0539} & \textBF{0.0845} \\
& &                  & (0.0130) & (0.0493) & (0.0904) & & (0.0061) & (0.0254) & (0.0440) \\\\
AMISE & NFR & deriv$_1$ & 0.2895 & 0.0449 & 0.0236  & & 0.2492 & 0.0324 & 0.0159 \\
		&		&	& (0.0905) & (0.0268) & (0.0151) & & (0.0509) & (0.0127) & (0.0099)		\\\\
		& & deriv$_2$ & 0.5418 & 0.0965 & 0.0443 & & 0.5374 & 0.0821 & 0.0345 \\
		&		&	& (0.1086) & (0.0348) & (0.0209) & & (0.0652) & (0.0151) & (0.0089)		\\\\
		& & pca$_1$ & 0.3303 & 0.0517 & 0.0265 & & 0.3163 & 0.0440 & 0.0199 \\
		&		&	& (0.0911)	 & (0.0235) & (0.0154) & & (0.0561) & (0.0148) & (0.0083) \\\\
		& & pca$_2$ & 0.2420 & 0.0377 & 0.0208 & & 0.2209 & 0.0272 & 0.0135 \\
		&		&	& (0.0818) & (0.0214) & (0.0137) & & (0.0461) & (0.0116) & (0.0070)	\\\\
		& & pca$_3$ & 0.1417 & 0.0257 & 0.0166 & & 0.0917 & 0.0169 & 0.0105 \\
		&		&	& (0.0612) & (0.0153) & (0.0123) & & (0.0304) & (0.0085) & (0.0079) 	\\\\
		& FSIM 	&	& \textBF{0.0473} & \textBF{0.0204} & \textBF{0.0155} & & \textBF{0.0236} & \textBF{0.0105} & \textBF{0.0069}	\\
		&		&	& (0.0678) & (0.0241) & (0.0215) & & (0.0366) & (0.0156) & (0.0080)	\\\\		
AMSPE & NFR & deriv$_1$ & 0.5373 & 0.6166 & 0.6992 & & 0.4782 & 0.5375 & 0.5922 \\
	&	&	& (0.2866) & (0.3246) & (0.3541) &	& (0.1850) & (0.2001) & (0.2181)	\\\\
	&	& deriv$_2$ & 1.2013 & 1.2174 & 1.2760 & & 1.1666 & 1.1869 & 1.2311	\\
	&	&	& (0.4803) & (0.4961) & (0.4998) & & (0.3367) & (0.3056) & (0.3149) \\\\
	&	& pca$_1$ & 0.6455 & 0.7024 & 0.7541 & &  0.6468 & 0.6899 & 0.7072 	\\
	&	& & (0.3258) & (0.3490) & (0.3694) & & (0.2413) & (0.2666) & (0.2605)	\\\\
	&	& pca$_2$ & 0.4303 & 0.4989 & 0.5625 & & 0.3758 & 0.4310 & 0.4697  	\\
	&	&	& (0.2553) & (0.2632) & (0.2706) & & (0.1466) & (0.1778) & (0.2115) \\\\
	&	& pca$_3$ & 0.2001 & 0.3312 & 0.4059 & & 0.1332 & 0.2242 & 0.2826 	\\
	&	&	& (0.1053) & (0.1893) & (0.2070) & & (0.0576) & (0.0980) & (0.1236) \\\\
	&	FSIM & & \textBF{0.0389} & \textBF{0.1201} & \textBF{0.1931} & & \textBF{0.0216} & \textBF{0.0573} & \textBF{0.0881}  \\
	& 	&  & (0.0268) & (0.0711) & (0.1154) & & (0.0160) & (0.0366) & (0.0585)	\\\bottomrule
  \end{longtable}
\end{center}

When errors are simulated from a Gaussian density with the AR$_{\rho=0.8}(1)$ structure, Table~\ref{tab:22} presents the AMSE, AMISE, and AMSPE for the functional single index model and the nonparametric functional regression models. As the signal-to-noise ratio decreases (i.e., $\xi$ value increases), the AMSE and AMSPE increase whereas the AMISE decreases. The functional single index model again produces much more accurate estimation and prediction accuracies than any nonparametric functional regression in all criteria. Compared to the results under the iid error structure in Table~\ref{tab:2}, the AMSE, AMISE, and AMSPE increase under the AR(1) error structure.

\begin{center}
\tabcolsep 0.095in
\renewcommand{\arraystretch}{0.82}
\begin{longtable}{@{}lllccccccc@{}}
  \caption{\small Estimation accuracy of the regression function between the functional single index model and the nonparametric functional regression with different choices of semi-metrics, for a set of rough curves under the Gaussian error density with the AR$_{\rho=0.8}(1)$ error structure and three different signal-to-noise ratios $\xi$.}\label{tab:22}\\
  \toprule
  & & & \multicolumn{3}{c}{$n=60$}  & & \multicolumn{3}{c}{$n=120$} \\
  Error & Model & Semi-metric& $\xi=0.1$ & $\xi=0.5$ & $\xi=0.9$ & & $\xi=0.1$ & $\xi=0.5$ & $\xi=0.9$ \\\toprule
\endfirsthead
  & & & \multicolumn{3}{c}{$n=60$}  & & \multicolumn{3}{c}{$n=120$} \\
  Error & Model & Semi-metric& $\xi=0.1$ & $\xi=0.5$ & $\xi=0.9$ & & $\xi=0.1$ & $\xi=0.5$ & $\xi=0.9$ \\\toprule
\endhead
\hline \multicolumn{10}{r}{{Continued on next page}} \\
\endfoot
\endlastfoot
AMSE & NFR & deriv$_1$ & 0.5976 & 0.9800 & 1.3103 & & 0.5003 & 0.7505 & 0.9787  \\
	  &	      &			 & (0.1569) & (0.3896) & (0.6410) & & (0.0999) & (0.2871) & (0.4866) \\
\\
	&		& deriv$_2$ & 1.2289 & 1.5027 & 1.7263 & & 1.1682 & 1.4443 & 1.6759 \\
	&		&		& (0.2051) & (0.3022) & (0.4338) & & (0.1814) & (0.3366) & (0.4951) \\
\\
	&		& pca$_1$ & 0.6855 & 0.9946 & 1.2908 & & 0.6527 & 0.8383 & 1.0221 \\
	&		& 		& (0.1457) & (0.3522) & (0.6014) & & (0.1052) & (0.2596) & (0.4389) \\
\\	 	
	&		& pca$_2$ & 0.4763 & 0.8079 & 1.1312 & & 0.4120 & 0.6180 & 0.8129 \\
	&		&		& (0.1136) & (0.3547) & (0.6338) & & (0.0681) & (0.2348) & (0.4180) \\
\\
	&		& pca$_3$ & 0.2995 & 0.6598 & 0.9667 & & 0.2030 & 0.4712 & 0.6891 \\
	&		&		& (0.0888) & (0.3257) & (0.5741) & & (0.0562) & (0.2485) & (0.4368) \\
\\				  
	&	FSIM	&  & \textBF{0.0943} & \textBF{0.4000} & \textBF{0.6900} & & \textBF{0.0615} & \textBF{0.2535} & \textBF{0.4358} \\
	&		& 		& (0.0635) & (0.3117) & (0.5527) & & (0.0473) & (0.2324) & (0.4156) \\
\\
AMISE & NFR & deriv$_1$ & 0.3476 & 0.0986 & 0.0617 & & 0.3269 & 0.0952 & 0.0655 \\
	&		&		& (0.0855) & (0.0406) & (0.0307) & & (0.0707) & (0.0326) & (0.0250) \\
\\
	&		& deriv$_2$ & 0.5321 & 0.1129 & 0.0791 & & 0.5246 & 0.1309 & 0.0825 \\
	&		&		& (0.1549) & (0.0497) & (0.0462) & & (0.0851) & (0.0173) & (0.0166) \\
\\			
	&		& pca$_1$ & 0.3826 & 0.1039 & 0.0677 & & 0.3601 & 0.0997 & 0.0665 \\
	&		&		& (0.0866) & (0.0417) & (0.0307) & & (0.0682) & (0.0320) & (0.0255) \\
\\
	&		& pca$_2$ & 0.3238 & 0.0924 & 0.0636 & & 0.3142 & 0.0923 & 0.0634 \\
	&		& 		& (0.0942) & (0.0399) & (0.0295) & & (0.0646) & (0.0333) & (0.0256) \\
\\
	&		& pca$_3$ & 0.2569 & 0.0850 & 0.0596 & & 0.2387 & 0.0856 & 0.0607 \\
	&		&		& (0.0985) & (0.0413) & (0.0307) & & (0.0698) & (0.0343) & (0.0261) \\
\\
	&	FSIM	&  & \textBF{0.0462} & \textBF{0.0270} & \textBF{0.0166} & & \textBF{0.0324} & \textBF{0.0109} & \textBF{0.0079} \\
	&		&		& (0.0380) & (0.0347) & (0.0263) & & (0.0276) & (0.0107) & (0.0068) \\
\\
AMSPE	& NFR & deriv$_1$ & 0.6660 & 1.0368 & 1.3756 & & 0.5142 & 0.7978 & 1.0303 \\
		&		&		& (0.3930) & (0.5867) & (0.8217) & & (0.1669) & (0.3369) & (0.5394) \\
\\
		&		& deriv$_2$ & 1.1967 & 1.7377 & 2.0512 & & 1.0869 & 1.2863 & 1.4891 \\
		&		&		& (0.6706) & (0.8520) & (0.9850) & & (0.1631) & (0.2664) & (0.3946) \\
\\
		&		& pca$_1$ & 0.7273 & 1.0383 & 1.3266 & & 0.6924 & 0.8834 & 1.0714 \\
		&		&		& (0.4081) & (0.6878) & (0.9066) & & (0.2390) & (0.3420) & (0.5087) \\
\\
		&		& pca$_2$ & 0.4942 & 0.8139 & 1.1263 & & 0.4193 & 0.6361 & 0.8390 \\
		&		&		& (0.2533) & (0.4369) & (0.6768) & & (0.1517) & (0.2941) & (0.4860) \\
\\
		&		& pca$_3$ & 0.3000 & 0.6563 & 0.9445 & & 0.2028 & 0.4842 & 0.7079 \\
		&		&		& (0.1592) & (0.3831) & (0.6064) & & (0.0943) & (0.2877) & (0.4809) \\
\\
		& FSIM	& & \textBF{0.1045} & \textBF{0.4205} & \textBF{0.7224} & & \textBF{0.0681} & \textBF{0.2655} & \textBF{0.4552} \\
		&		&	& (0.0686) & (0.3426) & (0.6097) & & (0.0526) & (0.2425) & (0.4370) \\
	\bottomrule			
  \end{longtable}
\end{center}

\subsection{Sparse curves}\label{sec:sparse_curves}

\begin{figure}[!htbp]
\centering
\subfloat[$n=60$]
{\includegraphics[width=8.55cm]{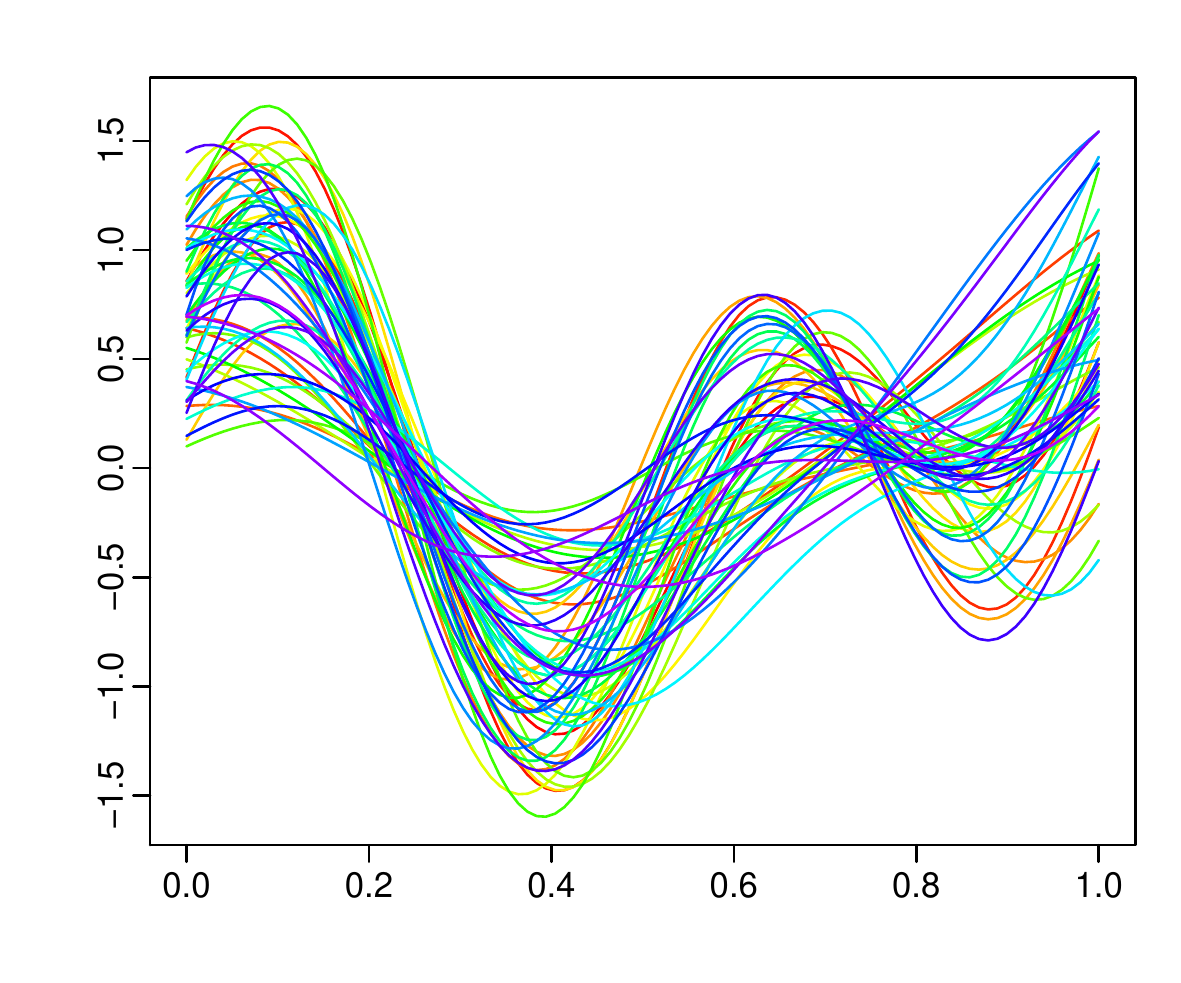}}
\qquad
\subfloat[$n=120$]
{\includegraphics[width=8.55cm]{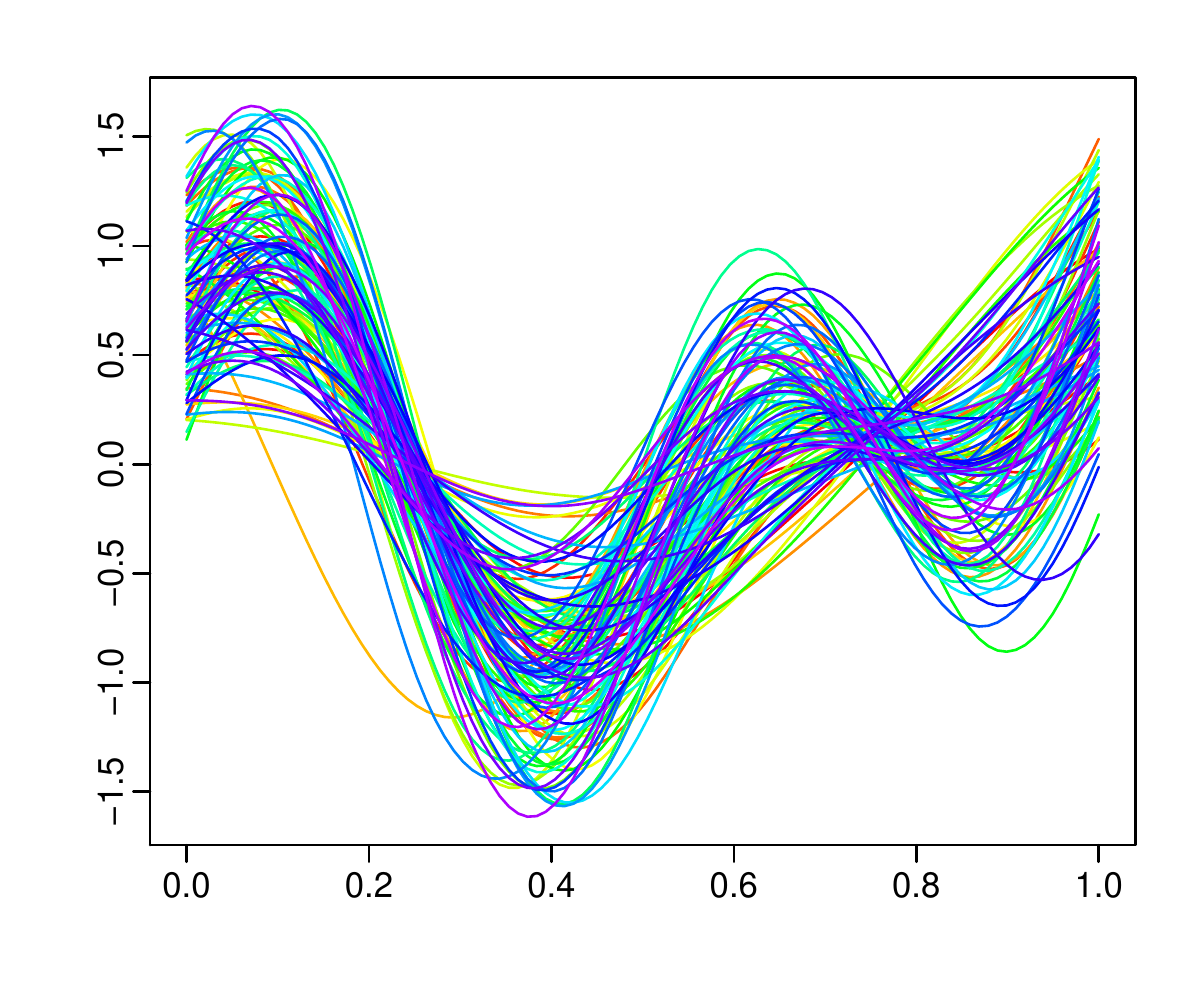}}
\caption{\small Simulated smooth sparse curves.}\label{fig:sparse_example}
\end{figure}

In the simulation study, we consider the same functional form as given in equation~\eqref{eq:smooth_curves}, but randomly select only 30 among 100 data points covering the support. The support for the sparse case is a subset of the original support. Figure~\ref{fig:sparse_example} presents the simulated curves for one replication with $n=60$ and $n=120$, respectively. 

Using the same setup as in Section~\ref{sec:smooth_curves}, Table~\ref{tab:22_sparse} and~\ref{tab:22_rough_sparse} present the AMSE, AMISE, and AMSPE for the functional single index model and the nonparametric functional regression models, under an iid Gaussian error density for the smoothed and rough curves. In the nonparametric functional regression, we study a range of semi-metrics, such as the semi-metric based on $1^{\textsuperscript{th}}$ and $2^{\textsuperscript{nd}}$ derivatives and the semi-metric based on the functional principal components. As the signal-to-noise ratio decreases (i.e., $\xi$ value increases), the AMSE and AMSPE increase, whereas the AMISE decreases. The functional single index model produces much more accurate estimation and prediction accuracies than any nonparametric functional regression in all criteria.  

\begin{center}
\tabcolsep 0.095in
\renewcommand{\arraystretch}{0.82}
\begin{longtable}{@{}lllccccccc@{}}
  \caption{\small Estimation accuracy of the regression function between the functional single index model and the nonparametric functional regression with different choices of semi-metrics, for a set of smooth and sparse curves under the Gaussian error density and three different signal-to-noise ratios $\xi$.}\label{tab:22_sparse}\\
  \toprule
  & & & \multicolumn{3}{c}{$n=60$}  & & \multicolumn{3}{c}{$n=120$} \\
  Error & Model & Semi-metric& $\xi=0.1$ & $\xi=0.5$ & $\xi=0.9$ & & $\xi=0.1$ & $\xi=0.5$ & $\xi=0.9$ \\\toprule
\endfirsthead
  & & & \multicolumn{3}{c}{$n=60$}  & & \multicolumn{3}{c}{$n=120$} \\
  Error & Model & Semi-metric& $\xi=0.1$ & $\xi=0.5$ & $\xi=0.9$ & & $\xi=0.1$ & $\xi=0.5$ & $\xi=0.9$ \\\toprule
\endhead
\hline \multicolumn{10}{r}{{Continued on next page}} \\
\endfoot
\endlastfoot
AMSE & NFR & deriv$_1$ & 0.8371 & 0.9224 & 0.9852 & & 0.5895 & 0.6534 & 0.6880 \\
	  &	      &			 & (0.3805) & (0.3956) & (0.3888) & & (0.2190) & (0.2769) & (0.2333) \\
\\
	&		& deriv$_2$ & 1.2054 & 1.2199 & 1.2342 & & 1.0067 & 1.0423 & 1.0806 \\
	&		&		    &  (0.2847) & (0.2921) & (0.2840) & & (0.2183) & (0.2207) & (1.2494) \\
\\
	&		& pca$_1$ & 0.6595 & 0.7152 & 0.7644 & & 0.6212 & 0.6509 & 0.6771 \\
	&		& 	 	 & (0.1497) & (0.1625) & (0.1770) & & (0.1078) & (0.1131) & (0.1240) \\
\\	 	
	&		& pca$_2$ & 0.3854 & 0.4519 & 0.5102 & & 0.3451 & 0.3825 & 0.4156 \\
	&		&		 & (0.0814) & (0.1176) & (0.1669) & & (0.0573) & (0.0642) & (0.0734) \\
\\
	&		& pca$_3$ & 0.3696 & 0.4466 & 0.5022 & & 0.3297 & 0.3705 & 0.3997 \\
	&		&		  & (0.0906) & (0.1234) & (0.1512) & & (0.0671) & (0.0739) & (0.0859) \\
\\				  
	&	FSIM	&  & \textBF{0.2295} & \textBF{0.2783} & \textBF{0.3274} & & \textBF{0.2070} & \textBF{0.2739} & \textBF{0.3021} \\
	&		& 		& (0.0634) & (0.0759) & (0.0978) & & (0.0497) & (0.0547) & (0.0623) \\
\\
AMISE & NFR & deriv$_1$ & 0.4092 & 0.0778 & 0.0395 & & 0.3194 & 0.0509 & 0.0239 \\
	&		&		& (0.1181) & (0.0385) & (0.0224) & & (0.1046) & (0.0280) & (0.0129) \\
\\
	&		& deriv$_2$ & 0.5629 & 0.1083 & 0.0487 & & 0.4829 & 0.0845 & 0.0385 \\
	&		&		& (0.0826) & (0.0352) & (0.0241) & & (0.0718) & (0.0222) & (0.0122) \\
\\			
	&		& pca$_1$ & 0.3505 & 0.0564 & 0.0290 & & 0.3207 & 0.0487 & 0.0228 \\
	&		&		& (0.0978) & (0.0257) & (0.0142) & & (0.0676) & (0.0171) & (0.0089) \\
\\
	&		& pca$_2$ & 0.2589 & 0.0400 & 0.0214 & & 0.2175 & 0.0307 & 0.0156 \\
	&		& 		& (0.0890) & (0.0222) & (0.0142) & & (0.0496) & (0.0131) & (0.0076) \\
\\
	&		& pca$_3$ & 0.2416 & 0.0405 & 0.0221 & & 0.2069 & 0.0294 & 0.0151 \\
	&		&		& (0.0858) & (0.0252) & (0.0150) & & (0.0513) & (0.0134) & (0.0068) \\
\\
	&	FSIM	&  & \textBF{0.1751} & \textBF{0.0239} & \textBF{0.0130} & & \textBF{0.1661} & \textBF{0.0201} & \textBF{0.0098} \\
	&		&		& (0.0888) & (0.0218) & (0.0155) & & (0.0470) & (0.0122) & (0.0070) \\
\\
AMSPE	& NFR & deriv$_1$ & 0.7660 & 0.8592 & 0.9073 & & 0.6595 & 0.7082 & 0.7360 \\
		&		&		& (0.4833) & (0.5310) & (0.5404) & & (0.3330) & (0.3796) & (0.3480) \\
\\
		&		& deriv$_2$ & 1.2089 & 1.1688 & 1.2654 & & 1.1496 & 1.1851 & 1.2500 \\
		&		&		& (0.5059) & (0.5622) & (0.5513) & & (0.3838) & (0.3960) & (0.3737) \\
\\
		&		& pca$_1$ & 0.6431 & 0.7016 & 0.7479 & & 0.6537 & 0.6934 & 0.7256 \\
		&		&		& (0.3066) & (0.3501) & (0.3808) & & (0.2159) & (0.2238) & (0.2365) \\
\\
		&		& pca$_2$ & 0.3967 & 0.4677 & 0.5231 &  & 0.3649 & 0.4142 & 0.4523 \\
		&		&		& (0.2335) & (0.3046) & (0.3797) & & (0.1541) & (0.1975) & (0.2246) \\
\\
		&		& pca$_3$ & 0.3842 & 0.4707 & 0.5115 & & 0.3574 & 0.4062 & 0.4369 \\
		&		&		& (0.2329) & (0.3056) & (0.3192) & & (0.1701) & (0.2059) & (0.2336) \\
\\
		& FSIM	& & \textBF{0.3214} & \textBF{0.3646} & \textBF{0.4049} & & \textBF{0.3031} & \textBF{0.3442} & \textBF{0.3789} \\
		&		&	& (0.1936) & (0.2209) & (0.2423) & & (0.1350) & (0.1555) & (0.1748) \\
	\bottomrule			
  \end{longtable}
\end{center}

\begin{center}
\tabcolsep 0.095in
\renewcommand{\arraystretch}{0.82}
\begin{longtable}{@{}lllccccccc@{}}
  \caption{\small Estimation accuracy of the regression function between the functional single index model and the nonparametric functional regression with different choices of semi-metrics, for a set of rough and sparse curves under the Gaussian error density and three different signal-to-noise ratios $\xi$.}\label{tab:22_rough_sparse}\\
  \toprule
  & & & \multicolumn{3}{c}{$n=60$}  & & \multicolumn{3}{c}{$n=120$} \\
  Error & Model & Semi-metric& $\xi=0.1$ & $\xi=0.5$ & $\xi=0.9$ & & $\xi=0.1$ & $\xi=0.5$ & $\xi=0.9$ \\\toprule
\endfirsthead
  & & & \multicolumn{3}{c}{$n=60$}  & & \multicolumn{3}{c}{$n=120$} \\
  Error & Model & Semi-metric& $\xi=0.1$ & $\xi=0.5$ & $\xi=0.9$ & & $\xi=0.1$ & $\xi=0.5$ & $\xi=0.9$ \\\toprule
\endhead
\hline \multicolumn{10}{r}{{Continued on next page}} \\
\endfoot
\endlastfoot
AMSE & NFR & deriv$_1$ & 1.0070 & 1.0501 & 0.1318 & & 0.9489 & 0.9991 & 1.0045 \\
	  &	      &			 & (0.2277) & (0.2418) & (0.3682) & & (0.1342) & (0.2034) & (0.1373) \\
\\
	&		& deriv$_2$ & 1.1528 & 1.1916 & 1.2917 & & 1.1342 & 1.1982 & 1.2025 \\
	&		&		    & (0.2727) & (0.2309) & (0.2579) & & (0.1380) & (0.1579) & (0.1660) \\
\\
	&		& pca$_1$ & 0.6544 & 0.7076 & 0.7567 & & 0.6472 & 0.6818 & 0.7139 \\
	&		& 	 	 & (0.1410) & (0.1550) & (0.1663) & & (0.1036) & (0.1093) & (0.1202) \\
\\	 	
	&		& pca$_2$ & 0.4319 & 0.5115 & 0.5905 & & 0.3845 & 0.4222 & 0.4551 \\	
	&		&		 & (0.1022) & (0.1521) & (0.2195) & & (0.0729) & (0.0788) & (0.0857) \\
\\
	&		& pca$_3$ & 0.4022 & 0.4760 & 0.5275 & & 0.3622 & 0.4074 & 0.4452 \\
	&		&		  & (0.1085) & (0.1577) & (0.1778) & & (0.0682) & (0.0805) & (0.0961) \\
\\				  
	&	FSIM	&      & \textBF{0.2505} & \textBF{0.3119} & \textBF{0.3716} & & \textBF{0.2707} & \textBF{0.2972} & \textBF{0.3265} \\
	&		& 		& (0.0712) & (0.0882) & (0.1149) & & (0.0516) & (0.0559) & (0.0642) \\
\\
AMISE & NFR & deriv$_1$ & 0.4649 & 0.0857 & 0.0429 & & 0.4466 & 0.0674 & 0.0286 \\
	&		&		& (0.1172) & (0.0345) & (0.0242) & & (0.0519) & (0.0144) & (0.0094) \\
\\
	&		& deriv$_2$ & 0.4893 & 0.0782 & 0.0340 & & 0.4753 & 0.0892 & 0.0372 \\
	&		&		& (0.1314) & (0.0324) & (0.0175) & & (0.0547) & (0.0207) & (0.0115) \\
\\			
	&		& pca$_1$ & 0.3416 & 0.0577 & 0.0300 & & 0.3348 & 0.0507 & 0.0235 \\
	&		&		& (0.0952) & (0.0265) & (0.0158) & & (0.0574) & (0.0168) & (0.0094) \\
\\
	&		& pca$_2$ & 0.2718 & 0.0459 & 0.0249 & & 0.2395 & 0.0334 & 0.0164 \\
	&		& 		& (0.0924) & (0.0251) & (0.0146) & & (0.0528) & (0.0134) & (0.0082) \\
\\
	&		& pca$_3$ & 0.2550 & 0.0407 & 0.0216 & & 0.2256 & 0.0317 & 0.0155 \\
	&		&		& (0.0879) & (0.0212) & (0.0130) & & (0.0489) & (0.0124) & (0.0075) \\
\\
	&	FSIM	&  & \textBF{0.1716} & \textBF{0.0285} & \textBF{0.0155} & & \textBF{0.1843} & \textBF{0.0213} & \textBF{0.0105} \\
	&		&		& (0.0678) & (0.0286) & (0.0180) & & (0.0457) & (0.0114) & (0.0071) \\
\\
AMSPE	& NFR & deriv$_1$ & 1.0655 & 1.1281 & 1.2161 & & 0.8677 &  0.9749 & 0.9983 \\
		&		&		& (0.5873) & (0.5264) & (0.5199) & & (0.2675) & (0.4007) & (0.2917) \\
\\
		&		& deriv$_2$ & 1.2579 & 1.2675 & 1.2723 & & 1.0808 & 1.1639 & 1.1859 \\
		&		&		   & (0.2607) & (0.3063) & (0.3502) & & (0.4120) & (0.3875) & (0.3943) \\
\\
		&		& pca$_1$ & 0.7011 & 0.7360 & 0.7755 & & 0.7004 & 0.7454 & 0.7807 \\
		&		&		  & (0.4045) & (0.4036) & (0.4221) & & (0.2699) & (0.3034) & (0.3161) \\
\\
		&		& pca$_2$ & 0.4421 & 0.5242 & 0.5868 & & 0.3936 & 0.4294 & 0.4615 \\
		&		&		 & (0.2854) & (0.4020) & (0.4772) & & (0.1492) & (0.1725) & (0.1908) \\
\\
		&		& pca$_3$ & 0.4017 & 0.4531 & 0.4966 & & 0.3690 & 0.4189 & 0.4583 \\
		&		&		 & (0.2319) & (0.2710) & (0.3024) & & (0.1598) & (0.1860) & (0.2209) \\
\\
		& FSIM	& 	& \textBF{0.3319} & \textBF{0.3964} & \textBF{0.4520} & & \textBF{0.3381} & \textBF{0.3659} & \textBF{0.3938} \\
		&		&	& (0.2111) & (0.2844) & (0.3367) & & (0.1470) & (0.1725) & (0.1889) \\
	\bottomrule			
  \end{longtable}
\end{center}

\subsection{Diagnostic check of Markov chains}\label{sec:3.4}

\begin{figure}[!htbp]
\centering
\includegraphics[width=8.31cm]{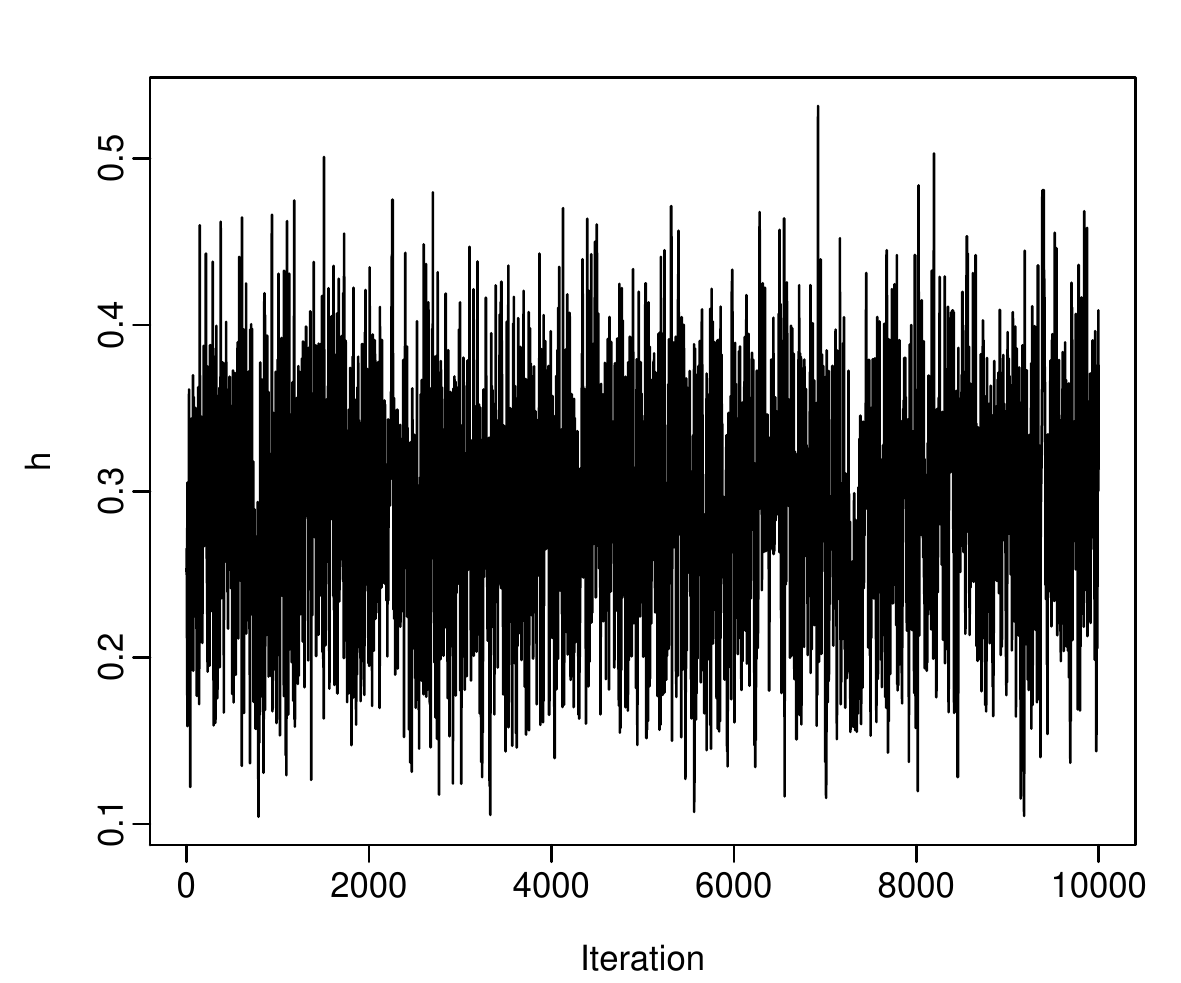}
\qquad
\includegraphics[width=8.31cm]{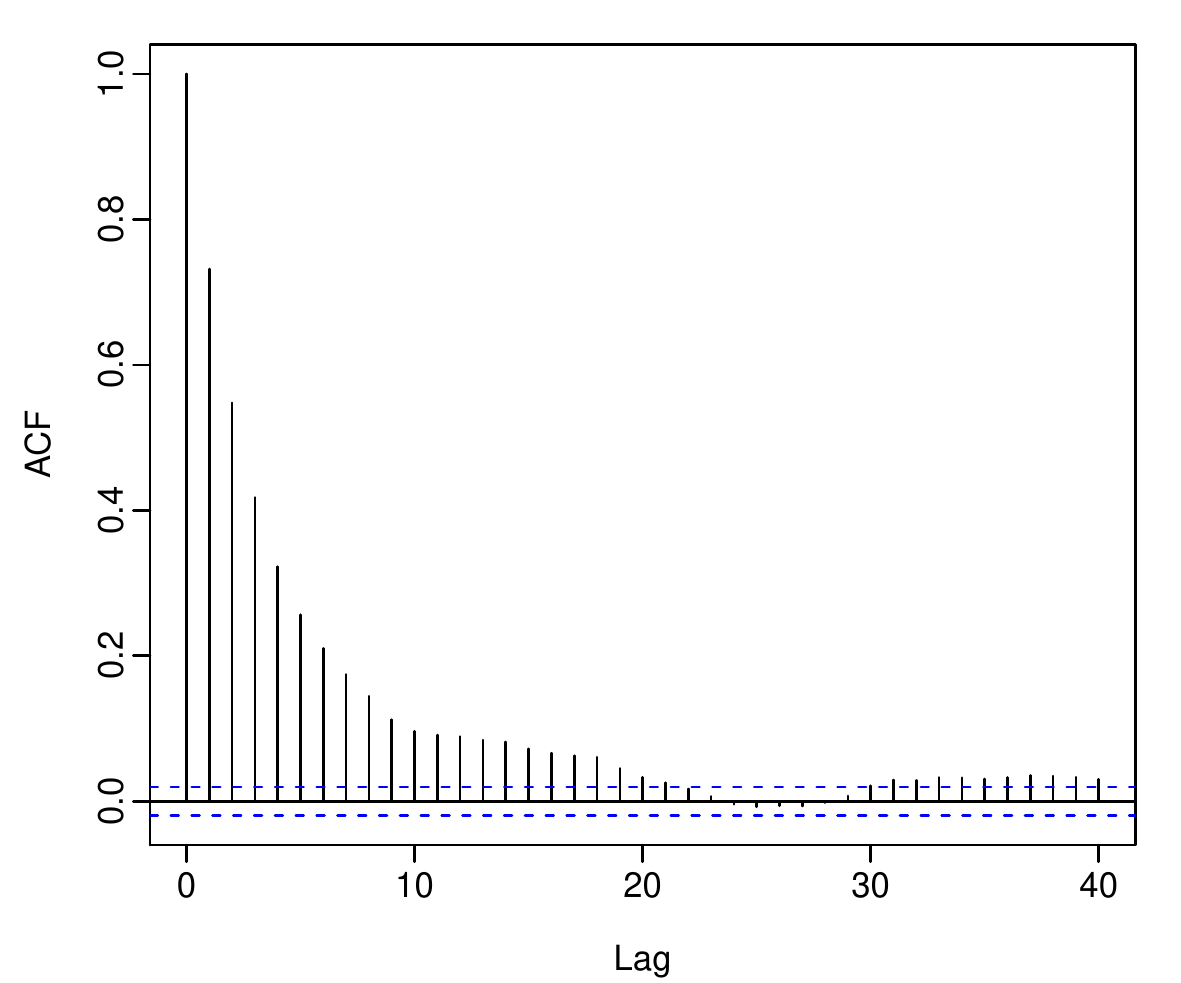}\\
\includegraphics[width=8.31cm]{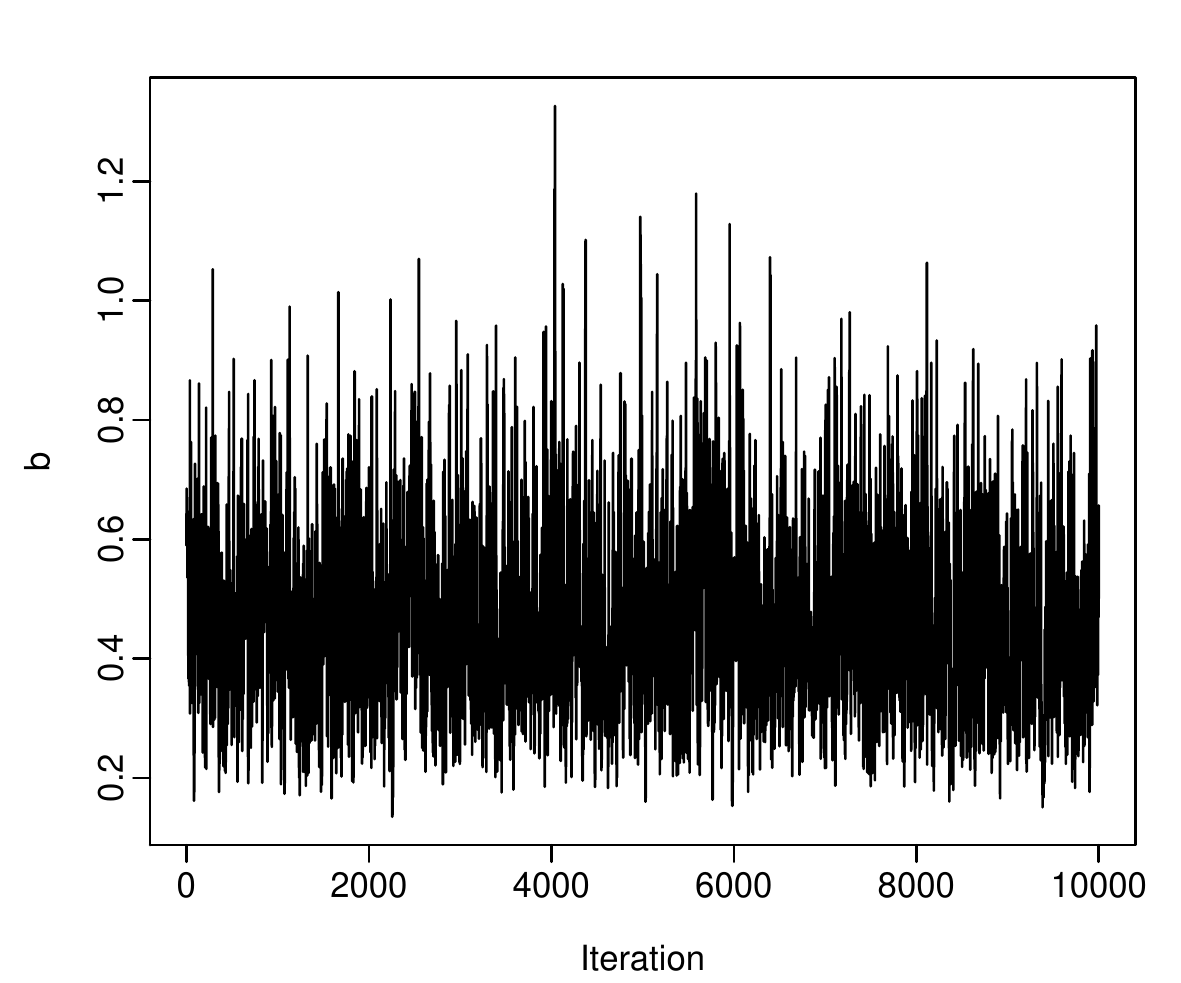} 
\qquad
\includegraphics[width=8.31cm]{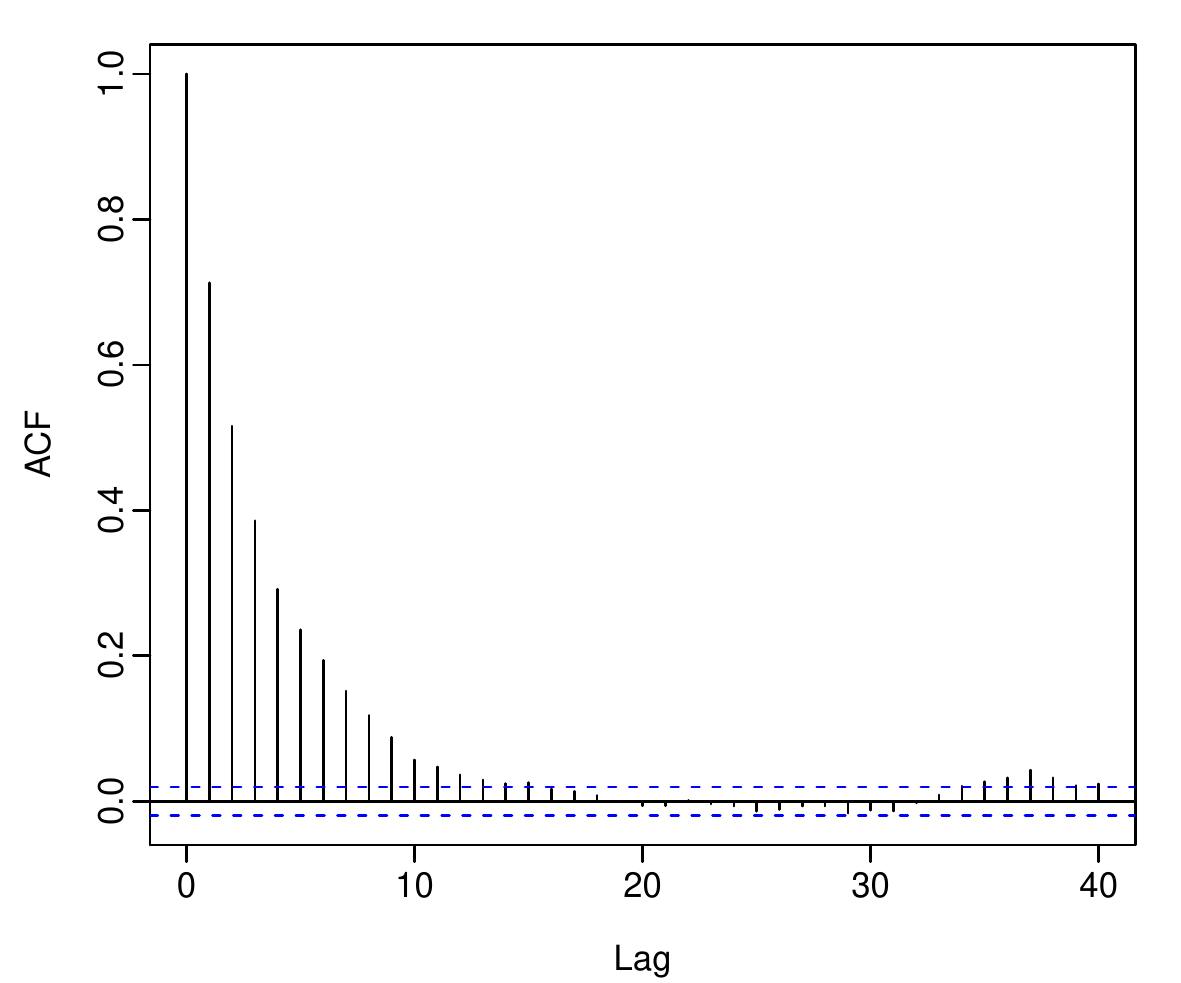}\\
\includegraphics[width=8.31cm]{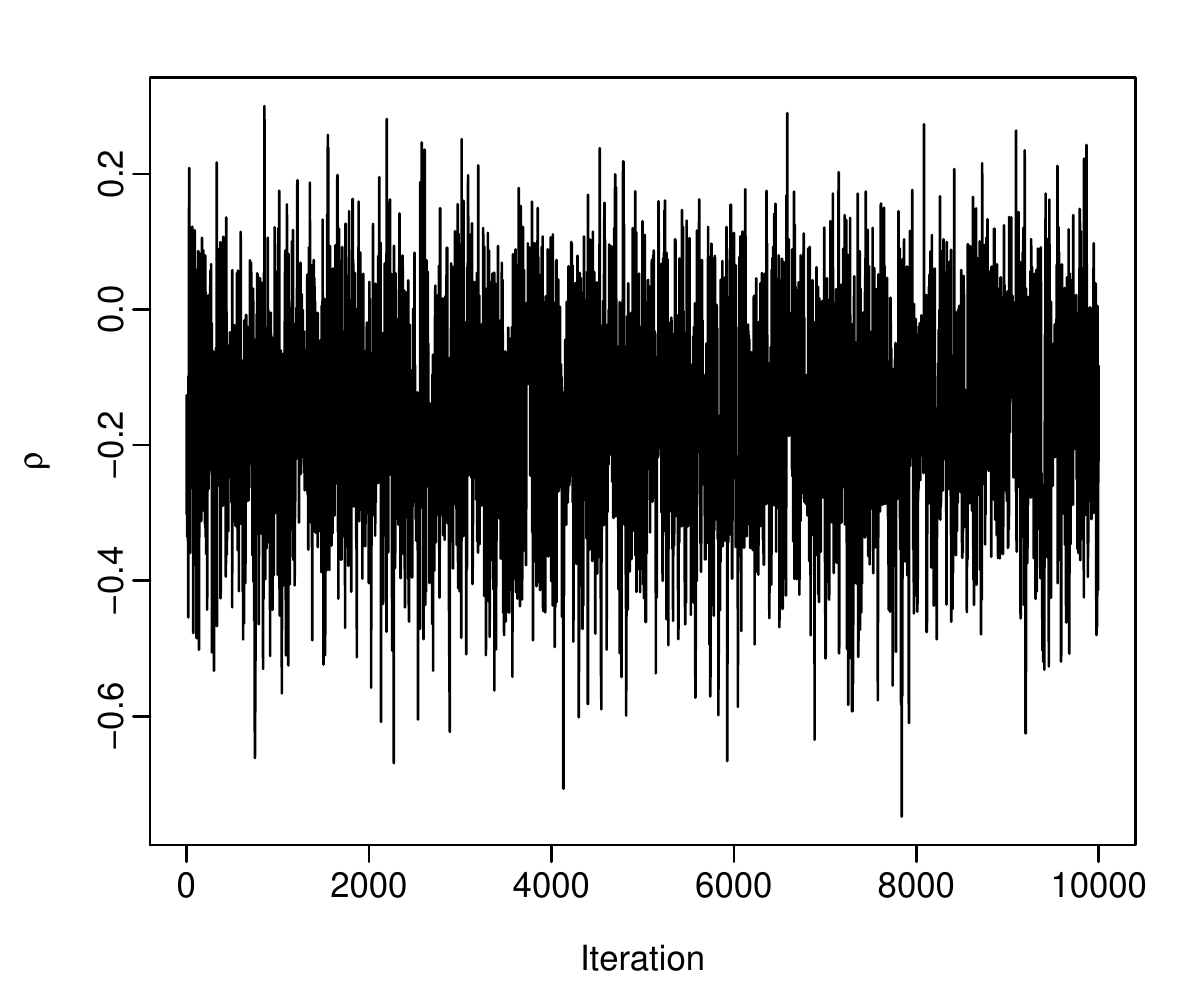}
\qquad
\includegraphics[width=8.31cm]{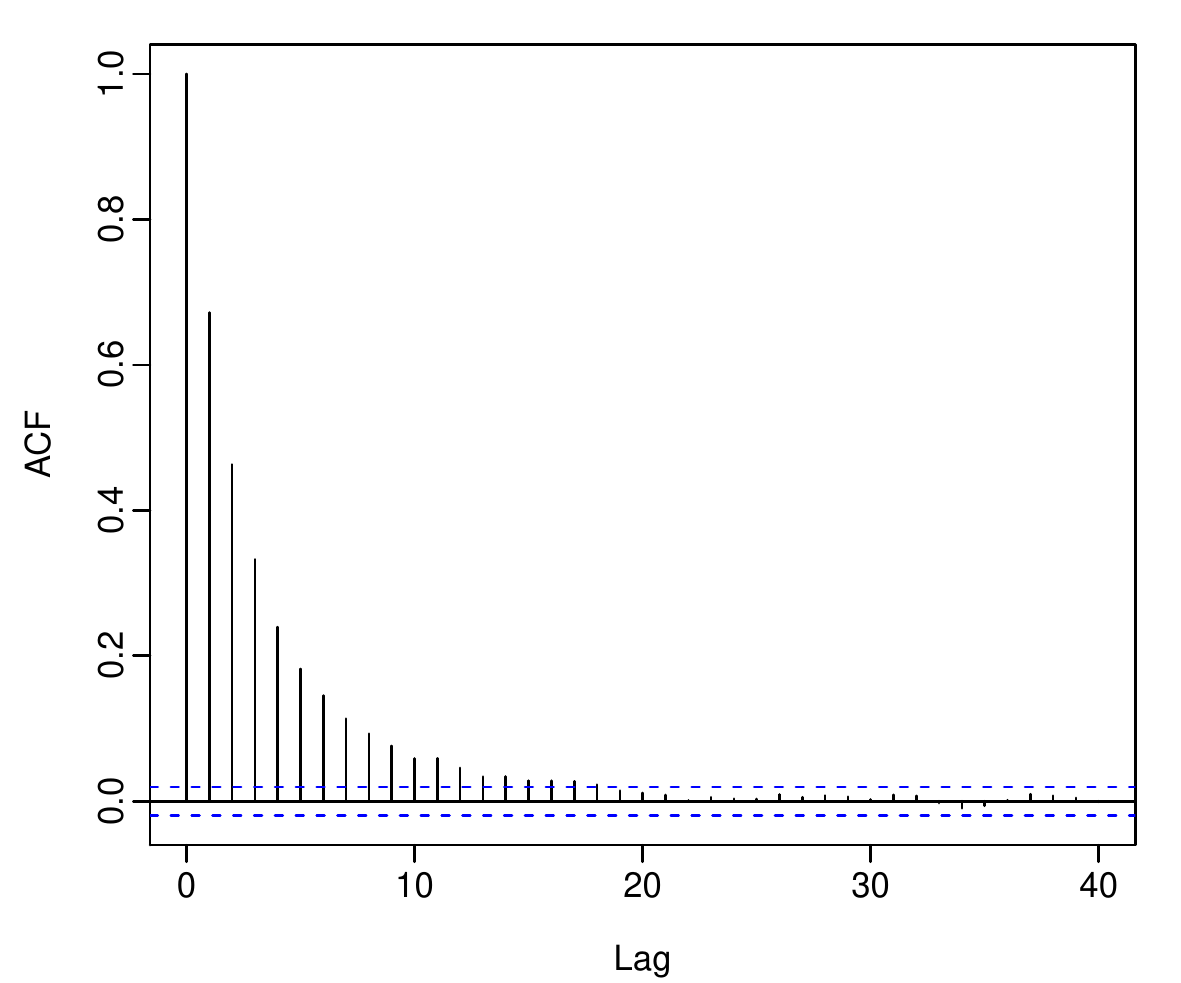}
\caption{\small MCMC sample paths and ACF of the sample paths, with Gaussian error density, the signal-to-noise ratio of 0.1, the sample size of 60 and the AR$_{\rho=0.8}(1)$ error structure.}\label{fig:diagnostic}
\end{figure}

As a demonstration with one replication, we plot the MCMC sample paths of the parameters in the left panel of Figure~\ref{fig:diagnostic}, and the autocorrelation functions of these sample paths in the right panel of Figure~\ref{fig:diagnostic}. These plots show that the sample paths are mixed well. Table~\ref{tab:ergodic} summarizes the ergodic averages, 95\% Bayesian credible intervals (CIs), sample SE, batch mean SE and SIF values. 

\begin{table}[!htbp]
\centering
\tabcolsep 0.25in
\caption{\small MCMC results of the bandwidth parameter estimation under the prior density of IG$(\alpha=1, \beta=0.05)$, with Gaussian error density, signal-to-noise ratio of 0.1, sample size of 60 and the AR$_{\rho=0.8}(1)$ error structure.}\label{tab:ergodic}
\begin{tabular}{@{}lrrlll@{}}
\toprule
\multicolumn{3}{l}{\hspace{-.267in}{Prior density: IG$(\alpha=1, \beta=0.05)$}} & \\\cmidrule{1-3}
Parameter & Mean & Bayesian CIs & SE & Batch mean SE & SIF \\\midrule
$h_n$ & 0.2917 & (0.1618, 0.4199) & 0.0677 & 0.1976 & 8.5317 \\
$b_n$ & 0.4641 & (0.2188, 0.8377) & 0.1598 & 0.4252 & 7.0771 \\
$\rho_n$ & -0.1564 & (-0.4695, 0.1215) & 0.1539 & 0.3668 & 5.6796  \\
\bottomrule
\end{tabular}
\end{table}

Using the coda package \citep{PBC+06}, we further checked the convergence of Markov chain with \citeauthor{Geweke92}'s \citeyearpar{Geweke92} convergence diagnostic test and \citeauthor{HW83}'s \citeyearpar{HW83} convergence diagnostic test. Our Markov chains pass both tests for all 100 replications.

\subsection{Analysis of sensitivity to prior choice}

To examine the robustness of the results with respect to the choice of the priors, we alter the prior densities in two ways. First, we keep the same prior distributions as before but alter the choice of hyper-parameters. Second, we change the prior distributions from an inverse Gamma distribution to a Cauchy distribution. The use of Cauchy prior to bandwidth estimation has been studied by \cite{ZBK09}. As summarized in Table~\ref{tab:sensitivity}, the MCMC results for the same set of samples are similar for the bandwidth parameters $h$ and $b$ but are different for the autocorrelation parameter $\rho$. 

\begin{table}[!htbp]
\centering
\tabcolsep 0.06in
\caption{\small MCMC results of the bandwidth parameter estimation under the different prior densities, with Gaussian error density, the signal-to-noise ratio of 0.1, the sample size of 60 and the AR$_{\rho=0.8}(1)$ error structure.}\label{tab:sensitivity}
\begin{tabular}{@{}lccccc@{}}
\toprule
Parameter & Mean & Bayesian CIs & SE & Batch mean SE & SIF \\\midrule
Prior density: IG$(\alpha=2, \beta=0.1)$ \\\cmidrule{1-1} 
$h_n$ & 0.2836 & (0.1642, 0.4141) & 0.0662 & 0.2014 & 9.2460  \\
$b_n$ & 0.4798 & (0.2097, 0.8466) & 0.1660 & 0.4460 & 7.2205 \\
$\rho_n$ & -0.2429 & (-0.5627, 0.0604) & 0.1549 & 0.3842 & 6.1535  \\
\midrule
Prior density: Cauchy$(x_0=0, \gamma=1)$ \\\cmidrule{1-1}
$h_n$ & 0.3259 & (0.2070, 0.4334) & 0.0582 & 0.1354 & 5.4116 \\
$b_n$ & 0.4304 & (0.2219, 0.7612) & 0.1388 & 0.3213 & 5.3596  \\
$\rho_n$ & -0.0243 & (-0.3021, 0.2081) & 0.1326 & 0.3481 & 6.8939 \\
\bottomrule
\end{tabular}
\end{table}

\section{Spectroscopy data}\label{sec:5}

We consider two near-infrared reflectances (NIR) spectroscopy data sets, which were previously studied by \cite{Kalivas97}, \cite{RO07}, and \cite{RO08}. These two datasets are available in the fds package \citep{SH13} in R \citep{Team17}.

\subsection{NIR spectra of wheat}

\begin{figure}[!htbp]
\centering
\includegraphics[width=11cm]{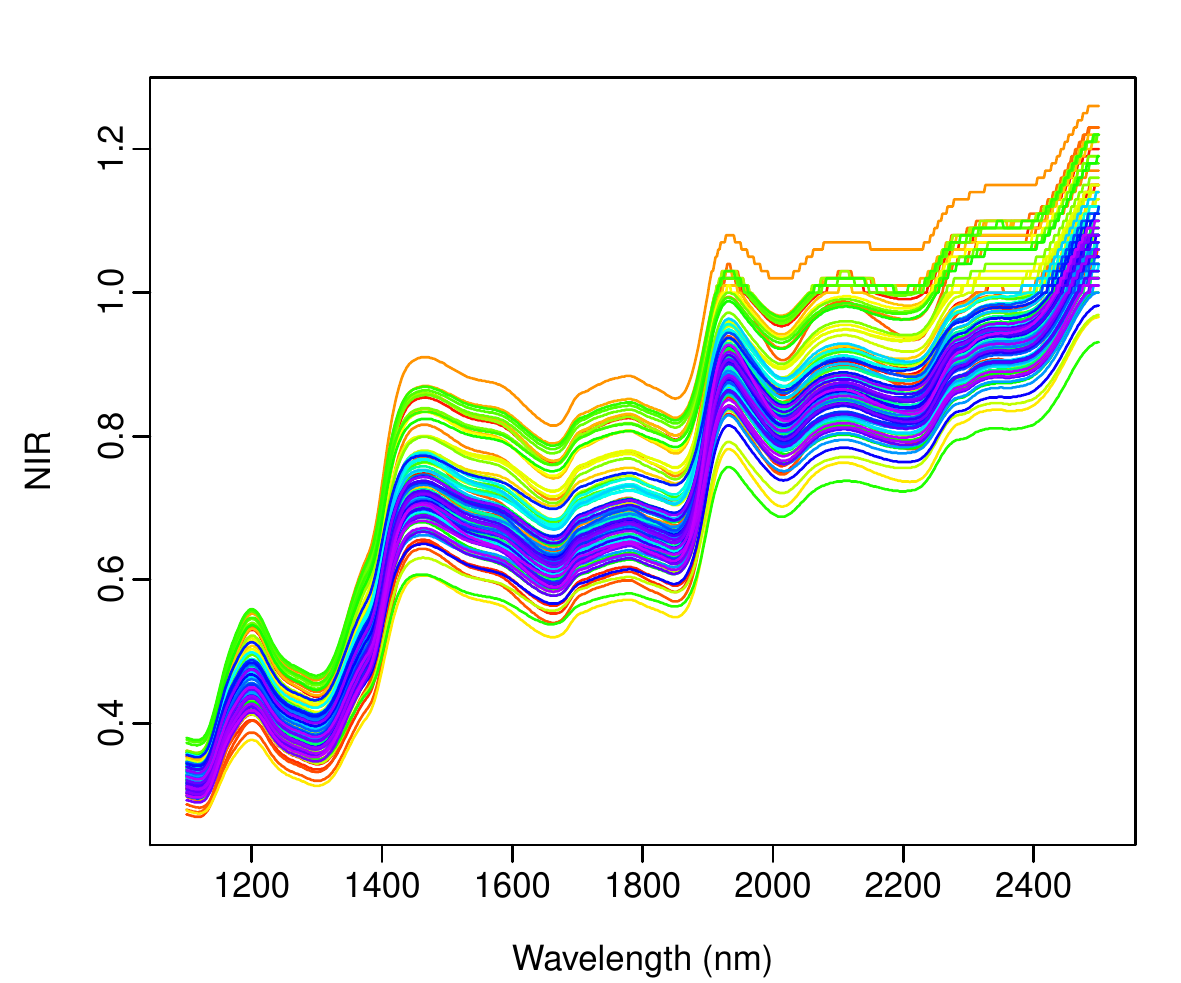}
\caption{\small A graphical display of the NIR spectra of 100 wheat samples.}\label{fig:NIR_moisture}
\end{figure}

The first data set consists of NIR spectra of 100 wheat samples, measured in 2nm intervals from 1100 to 2500nm, and an associated response variable (the samples' moisture contents). The ability to predict moisture in a wheat sample by the spectroscopic method has great practical value because high moisture content can lead to storage problems for wheat. A graphical display of the NIR spectra of wheat is presented in Figure~\ref{fig:NIR_moisture}.

We study the relationship between the spectrometric curves and the corresponding moisture content. We evaluate and compare the finite sample performance between the nonparametric functional regression with the NW estimator and the functional single index model. To assess the in-sample estimation accuracy and out-of-sample prediction accuracy of the regression models, we split the original 100 samples into two samples. The learning set contains 60 randomly selected samples, while the testing set contains the remaining 40 samples. The learning set allows us to build estimators of the regression function for both regression models, where the learning set is used. To measure the estimation and prediction accuracies, we evaluate and compare the forecast accuracy using the testing set, from which we predict responses in the testing set.

To measure the performance of each functional prediction method, we consider the MSE and MSPE. These are defined as
\begin{enumerate}
\item[(i)] MSE= $\frac{1}{60}\sum_{i=1}^{60}\left(y_i - \widehat{y}_i\right)^2$,
\item[(ii)] MSPE = $\frac{1}{40}\sum_{j=1}^{40}\left(y_{j} - \widehat{y}_{j}\right)^2$,
\end{enumerate}
where $i$ denotes an index for a randomly selected curve in the training set; whereas $j$ denotes an index for a randomly selected curve in the testing set.

Criterion (i) gives an indication of how well each in-sample observation is estimated, and the criterion (ii) measures how well each holdout observation is predicted. Averaged over 100 replications, the two different models and their corresponding values of MSE and MSPE are shown in Table~\ref{tab:moisture_mse_mspe}. There is an improvement in estimation and prediction accuracies for the functional single index model in comparison to the nonparametric functional regression. 

\begin{table}[!htbp]
\tabcolsep 0.33in
\renewcommand{\arraystretch}{0.62}
\centering
\caption{\small Estimation and prediction accuracies of the regression function between the functional single index model and the nonparametric functional regression with different choices of semi-metrics. For each of 100 replications, the training samples and testing samples were randomly shuffled.}\label{tab:moisture_mse_mspe}
\begin{tabular}{@{}lcccccc@{}}\toprule
& \multicolumn{5}{c}{NFR} & FSIM \\
& deriv$_1$ & deriv$_2$ & pca$_1$ & pca$_2$ & pca$_3$ &  \\\midrule
MSE & 1.81 &  0.72 &  1.40  & 	1.25 &  1.21	& \textBF{0.09} \\
	& (0.15) & (0.50) & (0.36)	& (0.33)	& (0.33)	& (0.04) \\\\
MSPE & 1.82 & 0.78 & 1.43	& 1.30	& 1.27	& \textBF{0.15} \\
	& (0.21) & (0.49) & (0.32)	& (0.33)	& (0.31)	& (0.08)  	\\\bottomrule	
\end{tabular}
\end{table}

\begin{figure}[!htbp]
\centering
\includegraphics[width=13.4cm]{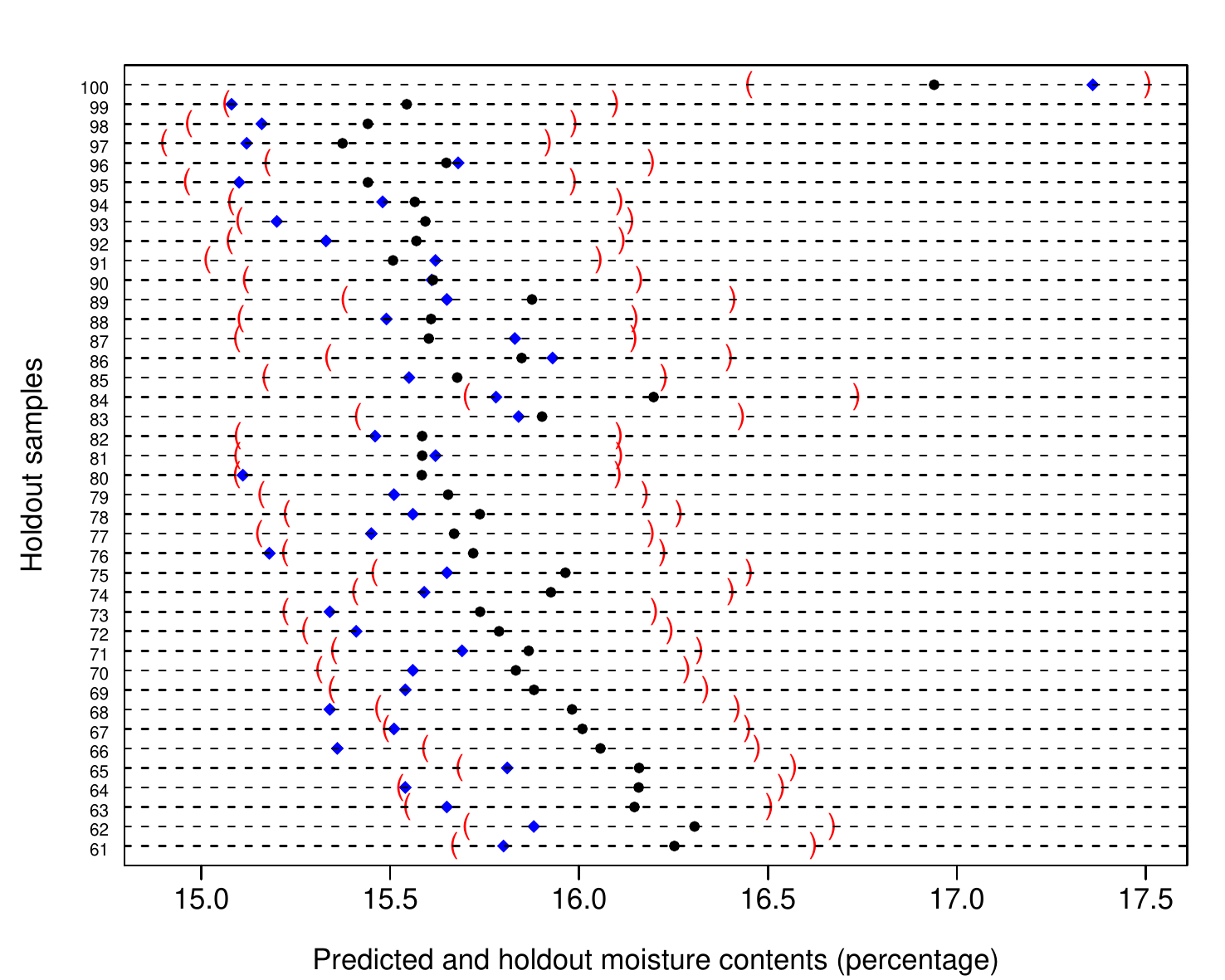}
\caption{\small The plot of predicted moisture contents via the functional single index model and the 95\% prediction intervals. The point forecasts of the moisture content are shown as black dots, the holdout samples are shown as blue diamonds, while the 95\% prediction intervals are shown as red parentheses.}\label{fig:wheat_PI}
\end{figure}

We are also interested in computing the prediction intervals nonparametrically, see Figure~\ref{fig:wheat_PI}. To this end, we first compute the cumulative density function (CDF) of the error distribution over a set of grid points within a range, such as -5 and 5. We then take the inverse of the CDF and find two grid points that are closest to the 2.5\% and 97.5\% quantiles; the 95\% prediction intervals of the holdout samples are obtained by adding the two grid points to the point forecasts. At the  95\% nominal coverage probability, the empirical coverage probability is 92.5\%. 

\subsection{NIR spectra of gasoline}

\begin{figure}[!htbp]
\centering
\includegraphics[width=12cm]{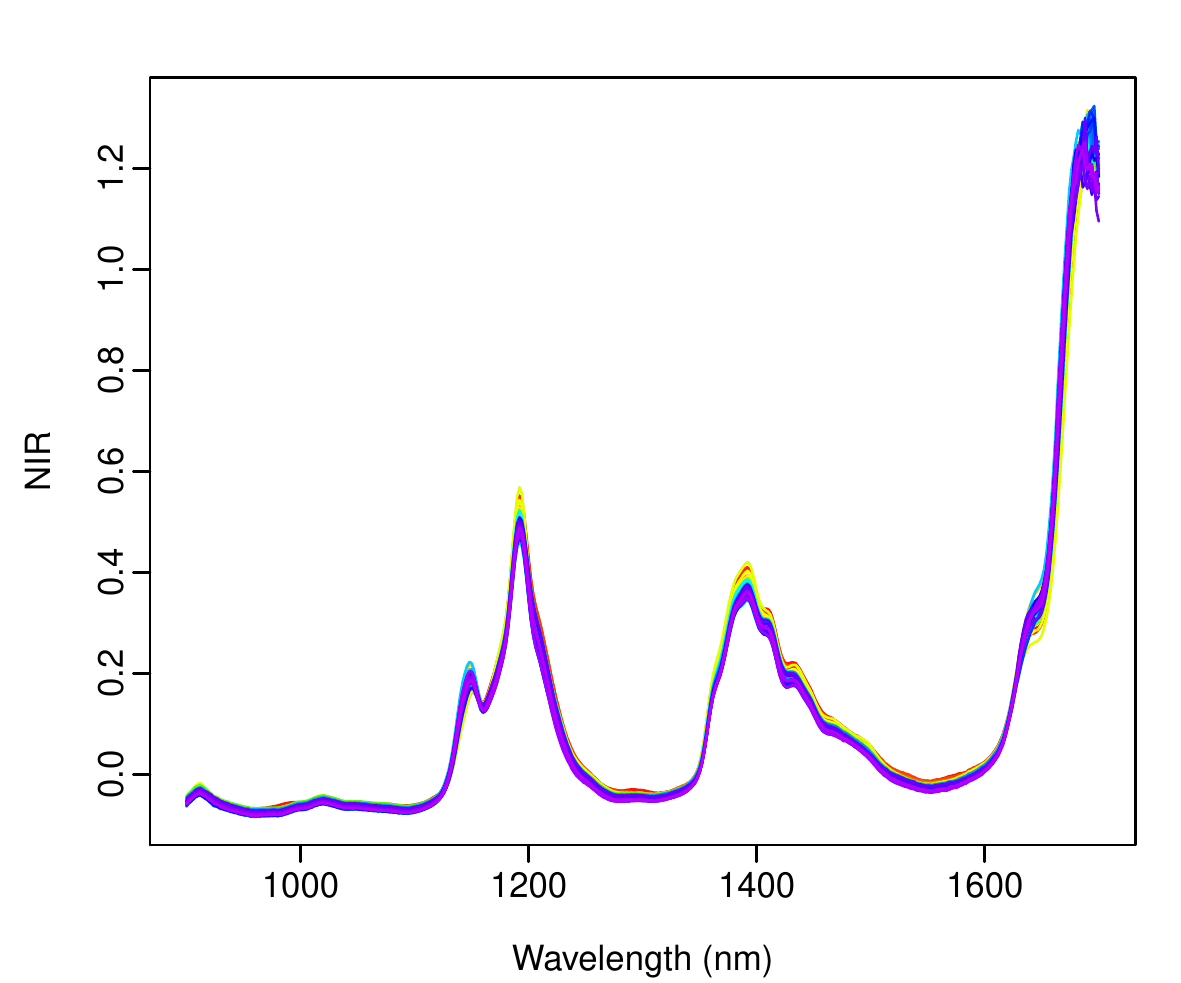}
\caption{\small A graphical display of the NIR spectra of 60 gasoline samples.}\label{fig:NIR}
\end{figure}

This data set contains 60 gasoline samples with specified octane numbers. Samples were measured using diffuse reflectance from 900 to 1,700nm in 2nm intervals, giving 401 wavelengths. The motivation is that obtaining a spectrometric curve is less time- and cost-consuming than the analytic chemistry needed for determining octane content. A graphical display of the NIR spectra of gasoline is presented in Figure~\ref{fig:NIR}.

The first step is to study the relationship between the spectrometric curves and the corresponding octane content, respectively. We evaluate and compare the finite sample performance between the nonparametric functional regression and functional single index model. To assess the in-sample estimation and out-of-sample forecast accuracies of the regression models, we split the original 60 samples into two subsamples \citep[see also][p. 105]{FV06}. The first one is called the learning sample, which contains 40 randomly selected samples. The second is called the testing sample, which contains the remaining 20 samples. The learning sample allows us to build the functional NW estimator with optimal bandwidth for both regression models, where the learning sample is used. To measure the estimation and prediction accuracies, we evaluate the functional NW estimator of the testing sample, from which we predict the responses in the testing sample.

Averaged over 100 replications, the two different models and their corresponding values of MSE and MSPE are shown in Table~\ref{tab:gas_mse_mspe}. Compared to the nonparametric functional regression, there is an improvement in estimation and prediction accuracies for the functional single index model.

\begin{table}[!htbp]
\tabcolsep 0.33in
\renewcommand{\arraystretch}{0.8}
\centering
\caption{\small Estimation and prediction accuracies of the regression function between the functional single index model and the nonparametric functional regression with different choices of semi-metrics. For each of 100 replications, the training samples and testing samples were randomly shuffled.}\label{tab:gas_mse_mspe}
\begin{tabular}{@{}lcccccc@{}}\toprule
& \multicolumn{5}{c}{NFR} & FSIM \\
& deriv$_1$ & deriv$_2$ & pca$_1$ & pca$_2$ & pca$_3$ & \\\midrule
MSE & 0.96 & 1.97 &  2.32 & 2.12 & 1.92 & \textBF{0.79} \\
	& (0.31) & (0.57) & (0.24) & (0.27) & (0.28) & (0.40) \\\\
MSPE & 1.49 & 2.08 & 2.46 & 2.29 & 2.12 & \textBF{1.41} \\
	& (0.62) & (0.69) &  (0.50) & (0.47) & (0.57) & (0.71) \\\bottomrule
\end{tabular}
\end{table}

\begin{figure}[!htbp]
\centering
\includegraphics[width=14.5cm]{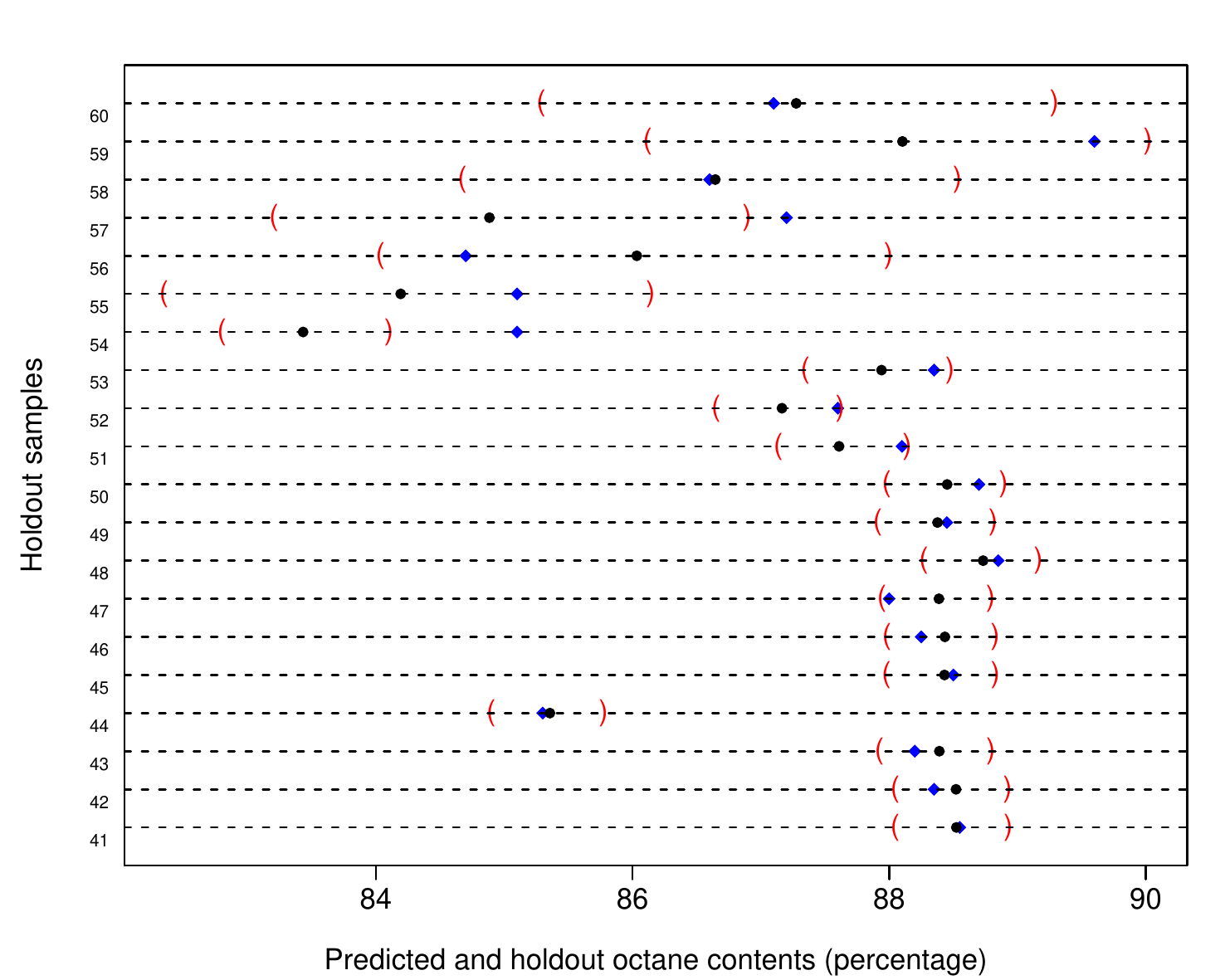}
\caption{\small The plot of predicted octane contents via the functional single index model and the 95\% prediction intervals. The point forecasts of the octane content are shown as black dots, the holdout samples are shown as blue diamonds, while the 95\% prediction intervals are shown as red parentheses.}\label{fig:Gas_PI}
\end{figure}

In Figure~\ref{fig:Gas_PI}, we display the 95\% pointwise prediction intervals for the predicted octane contents. At the  95\% nominal coverage probability, the empirical coverage probability is 90\%. 

\section{Conclusion}\label{sec:6}

We propose a Bayesian method to select optimal bandwidths in a functional single index model with possibly correlated errors and unknown error density. Through a series of simulations, the functional single index model outperforms the benchmark nonparametric functional regression, because the former is a semi-parametric model that selects the semi-metric in a data-driven manner. Using two spectroscopy data sets, the functional single index model produces much more accurate estimations and predictions than the nonparametric functional regression. The Bayesian bandwidth estimation approach allows the construction of nonparametric prediction interval for measuring the prediction uncertainty of the response variable.

There are many ways in which the proposed methodology can be extended, and we briefly mention a few at this point.
\begin{enumerate}
\item Consider other functional regression estimators, such as the functional local linear kernel estimator of \cite{BFR+07} or the $k$-nearest neighbor kernel estimator of \cite{BFV09}. The functional local linear kernel estimator can improve the estimation accuracy of the regression function by using a high-order kernel. The $k$-nearest neighbor kernel estimator takes into account the local structure of the data and gives better predictions when the functional data are heterogeneously concentrated.
\item Consider other bandwidth estimation methods for the kernel-form error density, such as the iterative methods proposed by \cite{MW90} and \cite{JMP91}, which are based on the relevant estimation of mean integrated square error.
\item Extend to functional single index model with heteroskedastic errors. Another kernel density estimator can estimate the covariate-dependent variance.
\item Extend to functional single index models with dependent functional data, where the functional predictors are the lagged values of the functional responses \citep[e.g.,][]{BCS00, QF11}.
\end{enumerate}

\newpage
\bibliographystyle{chicago}
\bibliography{single}

\begin{thebibliography}{}

\bibitem[\protect\citeauthoryear{{Ait-Sa\"{i}di}, Ferraty, Kassa, and
  Vieu}{{Ait-Sa\"{i}di} et~al.}{2008}]{AFK+08}
{Ait-Sa\"{i}di}, A., F.~Ferraty, R.~Kassa, and P.~Vieu (2008).
\newblock Cross-validated estimations in the single-functional index model.
\newblock {\em {Statistics: A Journal of Theoretical and Applied
  Statistics}\/}~{\em 42\/}(6), 475--494.

\bibitem[\protect\citeauthoryear{Akritas and {Van Keilegom}}{Akritas and {Van
  Keilegom}}{2001}]{AV01}
Akritas, M.~G. and I.~{Van Keilegom} (2001).
\newblock Non-parametric estimation of the residual distribution.
\newblock {\em Scandinavian Journal of Statistics\/}~{\em 28\/}(3), 549--567.

\bibitem[\protect\citeauthoryear{Benhenni, Ferraty, Rachdi, and Vieu}{Benhenni
  et~al.}{2007}]{BFR+07}
Benhenni, K., F.~Ferraty, M.~Rachdi, and P.~Vieu (2007).
\newblock Local smoothing regression with functional data.
\newblock {\em Computational Statistics\/}~{\em 22\/}(3), 353--369.

\bibitem[\protect\citeauthoryear{Besse, Cardot, and Stephenson}{Besse
  et~al.}{2000}]{BCS00}
Besse, P.~C., H.~Cardot, and D.~B. Stephenson (2000).
\newblock Autoregressive forecasting of some functional climatic variations.
\newblock {\em Scandinavian Journal of Statistics\/}~{\em 27\/}(4), 673--687.

\bibitem[\protect\citeauthoryear{Burba, Ferraty, and Vieu}{Burba
  et~al.}{2009}]{BFV09}
Burba, F., F.~Ferraty, and P.~Vieu (2009).
\newblock $k$-nearest neighbour method in functional nonparametric regression.
\newblock {\em Journal of Nonparametric Statistics\/}~{\em 21\/}(4), 453--469.

\bibitem[\protect\citeauthoryear{Chen, Hall, and M\"{u}ller}{Chen
  et~al.}{2011}]{CHM11}
Chen, D., P.~Hall, and H.-G. M\"{u}ller (2011).
\newblock Single and multiple index functional regression models with
  nonparametric link.
\newblock {\em The Annals of Statistics\/}~{\em 39\/}(3), 1720--1747.

\bibitem[\protect\citeauthoryear{Davis, Zang, and Zheng}{Davis
  et~al.}{2016}]{DZZ16}
Davis, R.~A., P.~Zang, and T.~Zheng (2016).
\newblock Sparse vector autoregressive modeling.
\newblock {\em Journal of Computational and Graphical Statistics\/}~{\em
  25\/}(4), 1077--1096.

\bibitem[\protect\citeauthoryear{Efromovich}{Efromovich}{2005}]{Efromovich05}
Efromovich, S. (2005).
\newblock Estimation of the density of regression errors.
\newblock {\em The Annals of Statistics\/}~{\em 33\/}(5), 2194--2227.

\bibitem[\protect\citeauthoryear{Escanciano and {Jacho-Ch\'{a}vez}}{Escanciano
  and {Jacho-Ch\'{a}vez}}{2012}]{EJ12}
Escanciano, J.~C. and D.~T. {Jacho-Ch\'{a}vez} (2012).
\newblock $\sqrt{n}$ uniformly consistent density estimation in nonparametric
  regression models.
\newblock {\em Journal of Econometrics\/}~{\em 167\/}(2), 305--316.

\bibitem[\protect\citeauthoryear{Fan and Gijbels}{Fan and Gijbels}{1996}]{FG96}
Fan, J. and I.~Gijbels (1996).
\newblock {\em {Local Polynomial Modelling and Its Applications}}.
\newblock London: Chapman \& Hall/CRC.

\bibitem[\protect\citeauthoryear{Fan, James, and Radchenko}{Fan
  et~al.}{2015}]{FJR15}
Fan, Y., G.~M. James, and P.~Radchenko (2015).
\newblock Functional additive regression.
\newblock {\em The Annals of Statistics\/}~{\em 43\/}(5), 2296--2325.

\bibitem[\protect\citeauthoryear{{Febrero-Bande}, Galeano, and
  {Gonz\'{a}lez-Manteiga}}{{Febrero-Bande} et~al.}{2017}]{FGG17}
{Febrero-Bande}, M., P.~Galeano, and W.~{Gonz\'{a}lez-Manteiga} (2017).
\newblock {Functional principal component regression and functional partial
  least-squares regression: An overview and a comparative study}.
\newblock {\em International Statistical Review\/}~{\em 85\/}(1), 61--83.

\bibitem[\protect\citeauthoryear{Ferraty, Laksaci, and Vieu}{Ferraty
  et~al.}{2005}]{FLV05}
Ferraty, F., A.~Laksaci, and P.~Vieu (2005).
\newblock Functional time series prediction via conditional mode estimation.
\newblock {\em Comptes Rendus Mathematique\/}~{\em 340\/}(5), 389--392.

\bibitem[\protect\citeauthoryear{Ferraty, Park, and Vieu}{Ferraty
  et~al.}{2011}]{FPV11}
Ferraty, F., J.~Park, and P.~Vieu (2011).
\newblock Estimation of a functional single index model.
\newblock In F.~Ferraty (Ed.), {\em {Recent Advances in Functional Data
  Analysis and Related Topics}}, Contributions to Statistics, pp.\  111--116.
  Springer.

\bibitem[\protect\citeauthoryear{Ferraty, Peuch, and Vieu}{Ferraty
  et~al.}{2003}]{FPV03}
Ferraty, F., A.~Peuch, and P.~Vieu (2003).
\newblock Single functional index model.
\newblock {\em Comptes Rendus Mathematique\/}~{\em 336\/}(12), 1025--1028.

\bibitem[\protect\citeauthoryear{Ferraty, Rabhi, and Vieu}{Ferraty
  et~al.}{2005}]{FRV05}
Ferraty, F., A.~Rabhi, and P.~Vieu (2005).
\newblock {Conditional quantiles for dependent functional data with application
  to the climatic El Ni\~{n}o phenomenon}.
\newblock {\em Sankhya: The Indian Journal of Statistics\/}~{\em 67\/}(2),
  378--398.

\bibitem[\protect\citeauthoryear{Ferraty and Vieu}{Ferraty and
  Vieu}{2006}]{FV06}
Ferraty, F. and P.~Vieu (2006).
\newblock {\em {Nonparametric Functional Data Analysis: Theory and Practice}}.
\newblock New York: Springer.

\bibitem[\protect\citeauthoryear{Garthwaite, Fan, and Sisson}{Garthwaite
  et~al.}{2016}]{GFS16}
Garthwaite, P.~H., Y.~Fan, and S.~A. Sisson (2016).
\newblock {Adaptive optimal scaling of Metropolis-Hastings algorithms using the
  Robbins-Monro process}.
\newblock {\em Communications in Statistics-Theory and Methods\/}~{\em
  45\/}(17), 5098--5111.

\bibitem[\protect\citeauthoryear{Geweke}{Geweke}{1992}]{Geweke92}
Geweke, J. (1992).
\newblock Evaluating the accuracy of sampling-based apapproach to calculating
  posterior moments.
\newblock In J.~M. Bernardo, J.~Berger, A.~P. Dawid, and J.~F.~M. Smith (Eds.),
  {\em Bayesian Statistics}, pp.\  169--193. Oxford: Clarendon Press.

\bibitem[\protect\citeauthoryear{Geweke}{Geweke}{2010}]{Geweke10}
Geweke, J. (2010).
\newblock {\em {Complete and Incomplete Econometric Models}}.
\newblock Princeton, USA: Princeton University Press.

\bibitem[\protect\citeauthoryear{Geweke}{Geweke}{1999}]{Geweke99}
Geweke, J.~F. (1999).
\newblock {Using simulation methods for Bayesian econometric models: Inference,
  development, and communication}.
\newblock {\em Econometric Reviews\/}~{\em 18\/}(1), 1--73.

\bibitem[\protect\citeauthoryear{Gilks, Richardson, and Spiegelhalter}{Gilks
  et~al.}{1996}]{GRS96}
Gilks, W.~R., S.~Richardson, and D.~J. Spiegelhalter (1996).
\newblock {\em {Markov chain Monte Carlo in Practice}}.
\newblock London: Chapman \& Hall.

\bibitem[\protect\citeauthoryear{Goia and Vieu}{Goia and Vieu}{2015}]{GV15}
Goia, A. and P.~Vieu (2015).
\newblock A partitioned single functional index model.
\newblock {\em Computational Statistics\/}~{\em 30\/}(3), 673--692.

\bibitem[\protect\citeauthoryear{Heidelberger and Welch}{Heidelberger and
  Welch}{1983}]{HW83}
Heidelberger, P. and P.~D. Welch (1983).
\newblock Simulation run length control in the presence of an initial
  transient.
\newblock {\em Operations Research\/}~{\em 31\/}(6), 1109--1144.

\bibitem[\protect\citeauthoryear{Hurvich and Tsai}{Hurvich and
  Tsai}{1989}]{HT89}
Hurvich, C.~M. and C.-L. Tsai (1989).
\newblock Regression and time series model selection in small samples.
\newblock {\em Biometrika\/}~{\em 76\/}(2), 297--307.

\bibitem[\protect\citeauthoryear{Hyndman and Shang}{Hyndman and
  Shang}{2010}]{HS10}
Hyndman, R.~J. and H.~L. Shang (2010).
\newblock {Rainbow plots, bagplots, and boxplots for functional data}.
\newblock {\em Journal of Computational and Graphical Statistics\/}~{\em
  19\/}(1), 29--45.

\bibitem[\protect\citeauthoryear{James and Silverman}{James and
  Silverman}{2005}]{JS05}
James, G.~M. and B.~W. Silverman (2005).
\newblock Functional adaptive model estimation.
\newblock {\em Journal of the American Statistical Association\/}~{\em
  100\/}(470), 565--576.

\bibitem[\protect\citeauthoryear{Jiang and Wang}{Jiang and Wang}{2011}]{JW11}
Jiang, C.-R. and J.-L. Wang (2011).
\newblock Functional single index models for longitudinal data.
\newblock {\em Annals of Statistics\/}~{\em 39\/}(1), 362--388.

\bibitem[\protect\citeauthoryear{Jones, Marron, and Park}{Jones
  et~al.}{1991}]{JMP91}
Jones, M.~C., J.~S. Marron, and B.~U. Park (1991).
\newblock A simple root-$n$ bandwidth selector.
\newblock {\em The Annals of Statistics\/}~{\em 19\/}(4), 1919--1932.

\bibitem[\protect\citeauthoryear{Kalivas}{Kalivas}{1997}]{Kalivas97}
Kalivas, J.~H. (1997).
\newblock Two data sets of near infrared spectra.
\newblock {\em Chemometrics and Intelligent Laborary Systems\/}~{\em 37\/}(2),
  255--259.

\bibitem[\protect\citeauthoryear{Kim, Shephard, and Chib}{Kim
  et~al.}{1998}]{KSC98}
Kim, S., N.~Shephard, and S.~Chib (1998).
\newblock {Stochastic volatility: Likelihood inference and comparison with ARCH
  models}.
\newblock {\em Review of Economic Studies\/}~{\em 65\/}(3), 361--393.

\bibitem[\protect\citeauthoryear{Marron and Wand}{Marron and Wand}{1992}]{MW92}
Marron, J.~S. and M.~P. Wand (1992).
\newblock {Exact mean integrated squared error}.
\newblock {\em Annals of Statistics\/}~{\em 20\/}(2), 712--736.

\bibitem[\protect\citeauthoryear{Meyer and Yu}{Meyer and Yu}{2000}]{MY00}
Meyer, R. and J.~Yu (2000).
\newblock {BUGS for a Bayesian analysis of stochastic volatility models}.
\newblock {\em Econometrics Journal\/}~{\em 3\/}(2), 198--215.

\bibitem[\protect\citeauthoryear{Morris}{Morris}{2015}]{Morris15}
Morris, J.~S. (2015).
\newblock Functional regression.
\newblock {\em Annual Review of Statistics and Its Application\/}~{\em 2},
  321--359.

\bibitem[\protect\citeauthoryear{M\"{u}ller and Wang}{M\"{u}ller and
  Wang}{1990}]{MW90}
M\"{u}ller, H.-G. and J.~L. Wang (1990).
\newblock Locally adaptive hazard smoothing.
\newblock {\em Probability Theory and Related Fields\/}~{\em 85\/}(4),
  523--538.

\bibitem[\protect\citeauthoryear{Neumeyer and Dette}{Neumeyer and
  Dette}{2007}]{ND07}
Neumeyer, N. and H.~Dette (2007).
\newblock Testing for symmetric error distribution in nonparametric regression
  models.
\newblock {\em Statistica Sinica\/}~{\em 17\/}(2), 775--795.

\bibitem[\protect\citeauthoryear{Plummer, Best, Cowles, and Vines}{Plummer
  et~al.}{2006}]{PBC+06}
Plummer, M., N.~Best, K.~Cowles, and K.~Vines (2006).
\newblock {CODA: Convergence diagnosis and output analysis for MCMC}.
\newblock {\em R News\/}~{\em 6\/}(1), 7--11.

\bibitem[\protect\citeauthoryear{{Quintela-del-R\'{i}o} and
  {Francisco-Fern\'{a}ndez}}{{Quintela-del-R\'{i}o} and
  {Francisco-Fern\'{a}ndez}}{2011}]{QF11}
{Quintela-del-R\'{i}o}, A. and M.~{Francisco-Fern\'{a}ndez} (2011).
\newblock {Nonparametric functional data estimation applied to ozone data:
  Prediction and extreme value analysis}.
\newblock {\em Chemosphere\/}~{\em 82\/}(6), 800--808.

\bibitem[\protect\citeauthoryear{{R Core Team}}{{R Core Team}}{2018}]{Team17}
{R Core Team} (2018).
\newblock {\em R: A Language and Environment for Statistical Computing}.
\newblock Vienna, Austria: R Foundation for Statistical Computing.
\newblock URL: \url{https://www.R-project.org/}.

\bibitem[\protect\citeauthoryear{Rachdi and Vieu}{Rachdi and Vieu}{2007}]{RV07}
Rachdi, M. and P.~Vieu (2007).
\newblock {Nonparametric regression for functional data: Automatic smoothing
  parameter selection}.
\newblock {\em Journal of Statistical Planning and Inference\/}~{\em 137\/}(9),
  2784--2801.

\bibitem[\protect\citeauthoryear{Reiss and Ogden}{Reiss and Ogden}{2008}]{RO08}
Reiss, P. and R.~T. Ogden (2008).
\newblock Smoothing parameter selection for a class of semiparametric linear
  models.
\newblock {\em Journal of the Royal Statistical Society: Series B\/}~{\em
  71\/}(2), 505--523.

\bibitem[\protect\citeauthoryear{Reiss, Goldsmith, Shang, and Ogden}{Reiss
  et~al.}{2017}]{RGS+17}
Reiss, P.~T., J.~Goldsmith, H.~L. Shang, and R.~T. Ogden (2017).
\newblock Methods for scalar-on-function regression.
\newblock {\em International Statistical Review\/}~{\em 85\/}(2), 228--249.

\bibitem[\protect\citeauthoryear{Reiss and Ogden}{Reiss and Ogden}{2007}]{RO07}
Reiss, P.~T. and R.~T. Ogden (2007).
\newblock Functional principal component regression and functional partial
  least squares.
\newblock {\em Journal of the American Statistical Association\/}~{\em
  102\/}(479), 984--996.

\bibitem[\protect\citeauthoryear{Robbins and Monro}{Robbins and
  Monro}{1951}]{RM51}
Robbins, H. and S.~Monro (1951).
\newblock A stochastic approximation method.
\newblock {\em The Annals of Mathematical Statistics\/}~{\em 22\/}(3),
  400--407.

\bibitem[\protect\citeauthoryear{Robert and Casella}{Robert and
  Casella}{2010}]{RC10}
Robert, C.~P. and G.~Casella (2010).
\newblock {\em {Introducing Monte Carlo Methods with R}}.
\newblock New York: Springer.

\bibitem[\protect\citeauthoryear{Roberts and Rosenthal}{Roberts and
  Rosenthal}{2009}]{RR09}
Roberts, G.~O. and J.~S. Rosenthal (2009).
\newblock {Examples of adaptive MCMC}.
\newblock {\em Journal of Computational and Graphical Statistics\/}~{\em
  18\/}(2), 349--367.

\bibitem[\protect\citeauthoryear{Shang}{Shang}{2013}]{Shang13}
Shang, H.~L. (2013).
\newblock Bayesian bandwidth estimation for a nonparametric functional
  regression model with unknown error density.
\newblock {\em Computational Statistics and Data Analysis\/}~{\em 67},
  185--198.

\bibitem[\protect\citeauthoryear{Shang}{Shang}{2014a}]{Shang14}
Shang, H.~L. (2014a).
\newblock {Bayesian bandwidth estimation for a functional nonparametric
  regression model with mixed types of regressors and unknown error density}.
\newblock {\em Journal of Nonparametric Statistics\/}~{\em 26\/}(3), 599--615.

\bibitem[\protect\citeauthoryear{Shang}{Shang}{2014b}]{Shang14b}
Shang, H.~L. (2014b).
\newblock Bayesian bandwidth estimation for a semi-functional partial linear
  regression model with unknown error density.
\newblock {\em Computational Statistics\/}~{\em 29\/}(3-4), 829--848.

\bibitem[\protect\citeauthoryear{Shang}{Shang}{2016}]{Shang16}
Shang, H.~L. (2016).
\newblock {A Bayesian approach for determining the optimal semi-metric and
  bandwidth in scalar-on-funciton quantile regression with unknown error
  density and dependent functional data}.
\newblock {\em Journal of Multivariate Analysis\/}~{\em 146}, 95--104.

\bibitem[\protect\citeauthoryear{Shang and Hyndman}{Shang and
  Hyndman}{2013}]{SH13}
Shang, H.~L. and R.~J. Hyndman (2013).
\newblock {\em fds: Functional data sets}.
\newblock University of Southampton.
\newblock R package version 1.7. URL:
  \url{https://CRAN.R-project.org/package=fds}.

\bibitem[\protect\citeauthoryear{Zhang, Brooks, and King}{Zhang
  et~al.}{2009}]{ZBK09}
Zhang, X., R.~D. Brooks, and M.~L. King (2009).
\newblock {A Bayesian approach to bandwidth selection for multivariate kernel
  regression with an application to state-price density estimation}.
\newblock {\em Journal of Econometrics\/}~{\em 153\/}(1), 21--32.

\bibitem[\protect\citeauthoryear{Zhang, King, and Shang}{Zhang
  et~al.}{2014}]{ZKS14}
Zhang, X., M.~L. King, and H.~L. Shang (2014).
\newblock A sampling algorithm for bandwidth estimation in an nonparametric
  regression model with a flexible error density.
\newblock {\em Computational Statistics and Data Analysis\/}~{\em 78},
  218--234.

\end{thebibliography}

\end{document}